\documentclass[11pt]{article}
\usepackage[utf8x]{inputenc}
\pdfoutput=1
\usepackage{amssymb,amsmath,mathrsfs,enumerate}
\usepackage{graphicx,rotate,multicol}
\usepackage{float}
\usepackage{tocloft}
\usepackage{subfig,subfloat}
\usepackage{multirow}
\usepackage{slashed,xcolor}
\usepackage[margin=10pt,labelfont=bf]{caption}
\usepackage{cite}
\usepackage{soul}
\usepackage{setspace}
\usepackage{jheppub}
\usepackage[export]{adjustbox}

\makeatletter
\let\Hy@linktoc\Hy@linktoc@page
\makeatother

\usepackage{color}
\definecolor{ourcolor}{rgb}{0.7, 0.25, 0.05}
\usepackage{tikz,braket}
\long\def\rpl#1!!#2!!{\textcolor{red}{#1} \textcolor{blue}{#2}}

\definecolor{My_red}        {cmyk}{0.00,1.00,1.00,0.10}

\def\slash#1{\rlap/#1}

\let\hat=\widehat
\let\bar=\overline

\def \order(#1){{\mathcal O} \left(#1 \right)}

\allowdisplaybreaks

\title{Broad toplike vector quarks at LHC and HL-LHC}

\author{Sayan Dasgupta,}
\author{Rohan Pramanick and}
\author{Tirtha Sankar Ray}
\affiliation{Department of Physics, Indian Institute of Technology Kharagpur,
Kharagpur 721302, India}

\emailAdd{sayandg05@gmail.com}
\emailAdd{rohanpramanick25@gmail.com}
\emailAdd{tirthasankar.ray@gmail.com}

\abstract{Top like vector quarks arising from underlying strong sectors are expected to have large decay widths pushing them beyond the narrow width approximation. In this paper we consider a broad colored vector quark that strongly couples to an exotic pseudoscalar. We use the full 1PI resummed propagator for the exotic quark to recast the present LHC constraints ruling out masses below $\sim1.2~(1.1)$ TeV for width to mass ratio of $0.1~(0.6)$. We utilize machine learning techniques that are demonstratively more efficient than traditional cut based searches to present the reach of HL-LHC on the parameter space of this broad resonance. We find that at $3~ab^{-1}$ the HL-LHC has a discovery potential up to $1.6$ TeV dominated by the pair production channel. We study the feasibility of using machine learning techniques to analyze the broad resonance peaks expected from these exotic quarks at collider experiments like the LHC.}

\begin{document}
\maketitle

\section{Introduction}
Vectorlike quarks are motivated exotic states that arise in various extensions of the Standard Model (SM) \cite{Panizzi:2014dwa}. If these states have nontrivial color charges they can represent optimistic new physics framework that can be discovered at a hadronic collider experiment like the Large Hadron Collider (LHC). There is a vast literature dedicated to the collider phenomenology of colored vectorlike quarks \cite{Kim:2018mks,Kim:2019oyh,Moretti:2016gkr,Cacciapaglia:2018qep,Alhazmi:2018whk,Durieux:2018ekg,Cacciapaglia:2018lld,Yepes:2018dlw,Carvalho:2018jkq,Liu:2018hum,Barducci:2017xtw,Deandrea:2017rqp,Liu:2017sdg,Liu:2016jho,Matsedonskyi:2015dns,Barducci:2015vyf,Backovic:2015bca,Chala:2014mma,Basso:2014apa,Matsedonskyi:2014mna,Backovic:2014uma,Gripaios:2014pqa,Han:2014qia,Karabacak:2014nca,Andeen:2013zca,Azatov:2013hya,Banfi:2013yoa,Li:2013xba,Barcelo:2011wu,Barcelo:2011vk, Cacciapaglia:2021uqh, Deandrea:2021vje,Balaji:2020qjg,Balaji:2021lpr}. In this paper we focus on the collider phenomenology of broad toplike vector quarks that arise from a strongly coupled sector. We consider that this broad vectorlike quark predominantly decays to a singlet pseudoscalar assumed to be a part of an underlying strong sector. The large coupling between the vectorlike quark and the scalar leads to a considerable decay width of the vectorlike quark pushing it beyond the narrow width approximation (NWA). While this scenario may be embedded in composite Higgs models \cite{Contino:2010rs} our analysis remains agnostic to the underlying model.\\\\
It has been pointed out that a one particle irreducible (1PI) propagator can capture some of the features of a broad resonance better than the NWA \cite{Azatov:2015xqa,Dasgupta:2019yjm}. In this paper, by incorporating the corrected propagator we demonstrate that for decay width to mass ratio ranging between $[0.05-0.6]$ the present LHC limit is in the $[0.9-1.15]$ TeV scale.\\\\
We study the HL-LHC reach to probe the parameter space of the broad colored toplike vector quark in both the pair and single production channels. We identify that the pair production channel with a $t\bar{t}b\bar{b}b\bar{b}$ final state is the most optimistic one. We utilize both the traditional cut and count search technique as well as machine learning (ML) classifiers to optimize the search efficiency in this channel \cite{Adhikary:2020fqf, Konar:2021nkk, Dey:2020tfq}. While there exists a plethora of different ML paradigms \cite{Albertsson:2018maf, Schwartz:2021ftp}, in this work we confine to boosted decision tree approach \cite{Roe:2004na} utilizing the extreme gradient boosting algorithm (XGBoost \cite{2016}). Expectedly the ML technique leads to a more aggressive reach in future runs of the LHC at $[1.5-1.6]$ TeV for width to mass ratio in the $[0.05-0.6]$ range at $3~ab^{-1}$ integrated luminosity. In the event of a discovery extracting physical parameters like the mass of any exotic state from such a broad resonance peak is a challenge. We investigate the possibility of employing ML techniques in the analysis of such broad resonance peaks beyond the NWA.\\\\
The rest of the  paper is organized as follows. In Section \ref{sec:model} we introduce the phenomenological Lagrangian for a vectorlike quark and a singlet pseudoscalar. In Section \ref{current_bound} we recast the searches at Run II of LHC to constraint the parameter space of these exotic states. In Section \ref{sec:future_strategy} we study the reach of HL-LHC for this broad resonance by including ML techniques. In Section \ref{sec:prediction} we study the feasibility of an ML tool for the analysis broad resonances before concluding.

\section{Effective Lagrangian}\label{sec:model}

In this section we introduce the phenomenological Lagrangian involving the toplike vector quark and  a pseudoscalar. We introduce a new charge $2/3$ vectorlike colored fermion $U$ of mass $M$ and hypercharge $7/6$. The new state $U$ is expected to mix with the SM third generation up-type quark. The Lagrangian after electroweak symmetry breaking can be parametrized as \cite{Cacciapaglia:2011fx},
\begin{equation}
\mathcal{L}_{eff} \supset i\bar{U}\slashed{D}U+i\bar{q}_L\slashed{D}
q_L+i\bar{\hat{t}}_R\slashed{D}
\hat{t}_R+\bar{b}_R\slashed{D}
b_R - \big[ \hat{m}_t\bar{\hat{t}}_L\hat{t}_R+m_{mix}\bar{U}_L\hat{t}_R+M\bar{U}_LU_R + h.c.\big],  
\label{master:lgn}
\end{equation}
where, $q_L=\{\hat{t}_L,b_L\}^T$ and two singlets $\hat{t}_R$ and $b_R,$ are the usual third generation SM  quarks in the gauge basis. The left handed components of $U$ and $\hat{t}$ mix with an angle $\theta_L$ and the right handed components mix with an angle $\theta_R$ resulting in the mass basis states $t$ which is identified as the SM top and a new exotic resonance $t^{\prime}$.\\\\
The mixing angles are correlated and is given by
\begin{equation}
\begin{split}
&\sin\theta_R=\frac{M}{m_t}\sin\theta_L  \\
\mbox{where,}~~ \hspace{0.1in}&M^2=\frac{m_{t^{\prime}}^2+\sin^2\theta_Rm_t^2(m_{t^{\prime}}^2-m_t^2)}{1+\sin^2\theta_R(m_{t^{\prime}}^2-m_t^2)}
\end{split}
\label{eq:paramrelations}
\end{equation}
The relevant SM couplings of $t^{\prime}$ are listed in Table~\ref{tab:couplings}.
\begin{table}[t]
	\centering
	\begin{tabular}{ cc }
		\hline\hline
		Vertex & Coupling \\ 
		\hline\\
		$\bar{t^{\prime}}\slash{G}t^{\prime}$ & $g_s$ \\\\
		$\bar{t^{\prime}}\slash{Z}t$ & $\frac{1}{4}(\sin2\theta_R+2\sin2\theta_L)(g_W\cos\theta_W+g_1\sin\theta_W)$ \\\\
		$\bar{t^{\prime}}\slash{W}b$ & $\frac{1}{\sqrt{2}}g_W\sin\theta_L$ \\\\
		$\bar{t^{\prime}}_LHt_R$ & $\frac{\sqrt{2}}{v}\cos\theta_R(m_{mix}\cos\theta_L+\hat{m}_t\sin\theta_L)$ \\\\
		$\bar{t^{\prime}}_RHt_L$ & $\frac{\sqrt{2}}{v}\sin\theta_R(m_{mix}\sin\theta_L-\hat{m}_t\cos\theta_L)$ \\\\
		\hline\hline
	\end{tabular}
	\caption{\small\it Relevant SM couplings of $t^{\prime}$. Here $g_s$ is the strong coupling, $g_W$ is the weak coupling, $g_1$ is the $U(1)_Y$ coupling, $\theta_W$ is the weinberg angle and $v$ is the Higgs vacuum expectation value.}
	\label{tab:couplings}
\end{table}
The large decay width of $t^{\prime}$ is driven by its decay to a pseudoscalar $\phi$ with mass $M_{\phi}$. The effective Lagrangian can be written as \cite{Bizot:2018tds}
\begin{equation}
\mathcal{L}_{eff} \supset g_{\phi}^{\ast}\bar{t^{\prime}}\phi t-i\frac{C^{\phi}_tm_t}{f_{\phi}}\phi\bar{t}\gamma^5t-i\frac{C^{\phi}_bm_b}{f_{\phi}}\phi\bar{b}\gamma^5b+h.c.
\label{master:lgn:a}
\end{equation}
In models of composite Higgs where the Higgs is a pseudo-Nambu Goldstone boson (pNGB) of a strong sector \cite{Contino:2010rs} and gets its masses through partial compositeness, the vectorlike quark may be identified with a top partner \cite{DeSimone:2012fs}. 
The field $\phi$ represents a physical pseudoscalar state that is assumed to be a part of the same strong sector as the toplike top partner. Due to the assumed strong interaction the vectorlike top partner predominantly decays to this pseudoscalar state which results in the large decay width of this vectorlike quark. The pseudoscalar is assumed to predominantly decay to a pair of bottom quarks. While we introduce this pseudoscalar by hand in the low energy effective Lagrangian they commonly arise in a wide variety of composite Higgs models as a part of an extended Higgs sector \cite{Gripaios:2009pe,Mrazek:2011iu,Bertuzzo:2012ya}. Interestingly some of these extended scenarios are more natural in 4D UV completions of composite Higgs models \cite{Cacciapaglia:2019bqz,Ferretti:2014qta,Barnard:2013zea}. An extensive literature exists on the phenomenology of this exotic pseudoscalar \cite{Bizot:2018tds,Cacciapaglia:2019zmj,Chala:2017xgc}.\\\\
At the LHC the exotic state $t^{\prime}$ is dominantly pair produced through gluon fusion. Once they are produced they will dominantly decay through the channels: $\phi t$, $Ht, Zt$ and $Wb.$ In this paper we set $\sin\theta_R=0.1$ which is a conservative choice in consonance with electroweak precision constraints \cite{Cacciapaglia:2011fx}. Following \cite{Bizot:2018tds} we set $C^{\phi}_t$, $C^{\phi}_b$ and $f_{\phi}$ at $1.46$, $1.9$ and $2.8$ TeV respectively which are consistent with the minimal fundamental composite Higgs models with the $SU(4)/Sp(4)$ coset \cite{Ferretti:2014qta,Barnard:2013zea,Erdmenger:2020lvq}. We trade $g_{\phi}^{\ast}$ for the decay width of $t^{\prime}$ ($\Gamma_{t^{\prime}}/m_{t^{\prime}}$) which we identify as a free parameter for our analysis. The branching ratios of $t^{\prime}$ for representative values of $m_{t^{\prime}}$ are listed in Table~\ref{tab:branching_ratio}.
\begin{table}[t]
	\centering
	\begin{tabular}{ cccccc }
		\hline\hline
		$m_{t^{\prime}}$ (TeV) & $\Gamma_{t^{\prime}}/m_{t^{\prime}}[\%]$& $Br(t^{\prime}\rightarrow\phi t)$ & $Br(t^{\prime}\rightarrow Z t)$ & $Br(t^{\prime}\rightarrow H t)$ & $Br(t^{\prime}\rightarrow W b)$\\ 
		\hline
		$0.8$ & $5$ & $0.946$ & $0.021$ & $0.033$ & $-$ \\
		$1.3$ & $30$ & $0.974$ & $0.009$ & $0.017$ & $0.0003$\\
		\hline\hline
	\end{tabular}
	\caption{\small\it Branching ratios of $t^{\prime}$. Here $M_{\phi}=100$ GeV.}
	\label{tab:branching_ratio}
\end{table}
The pseudoscalar $\phi$ decays dominantly to a pair of bottom quarks when the decay to top quarks is not kinematically allowed. For $M_{\phi}>2m_t$, $\phi$ primarily decays to a pair of top quarks.

\subsection{Broad resonance and the full 1PI propagator}\label{beyond:nwa}

\begin{figure}[t]
	\centering
	\adjincludegraphics[trim={5cm 23cm 5cm 3cm},clip ,scale=0.6]{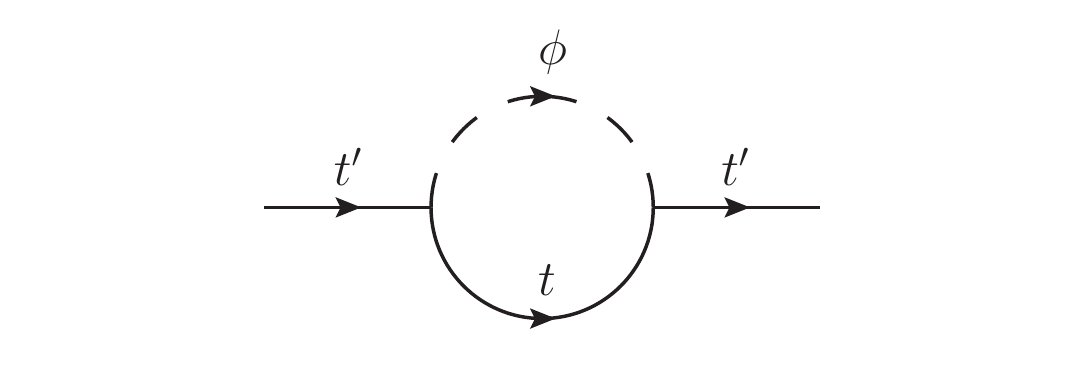}
	\caption{\small\it The primary loop contribution to the $t^{\prime}$ propagator.}
	\label{fig:top_partner_loop}
\end{figure}

It has been pointed out in \cite{Azatov:2015xqa} and \cite{Dasgupta:2019yjm} that the Breit Wigner (BW) form of a propagator may not be appropriate for broad resonances where the width to mass ratio is greater than $10 \%$. In our setup the colored vectorlike quark $t^{\prime}$ has a strong coupling to the pseudoscalar $\phi$. Owing to this large coupling, the decay width of $t^{\prime}$ becomes large enough to violate the NWA. To account for this we will use the full 1PI resummed propagator that captures the leading effects of the large width:
\begin{equation}
\mathcal{D}_{t^{\prime}}(p)=\frac{1}{\slash{p}-m_{t^{\prime}}+iIm(\Sigma)}
\label{eq:prop}
\end{equation}
where $\Sigma$ is the contribution from the loop shown in Figure~\ref{fig:top_partner_loop}. The imaginary part of $\Sigma$ is given by
\vspace{-0.75em}
\begin{eqnarray}
\dfrac{1}{2} \mathrm{Tr} (\slash{p}  + m_{t^\prime}) Im(\Sigma)=\frac{g_{\phi}^{\ast 2}}{16\pi}\theta(p - m_t - M_\phi) & \sqrt{(p^2-m_t^2-M_{\phi}^2)^2-4m_t^2M_{\phi}^2} \nonumber \\
& \times \Bigg[1+\dfrac{m_t^2-M_{\phi}^2+2m_{t^{\prime}}m_t}{p^2}\Bigg]
\label{eq:1pi_im}
\end{eqnarray}
The finite contribution from the real part of $\Sigma$ may be construed to make the top partner mass momentum dependent. However, as demonstrated in Appendix~\ref{appn:mom_dependence}, the momentum dependence is mild and leads to numerically insignificant variation in the effective mass of top partner within the resonance peak \cite{Liu:2019bua}. We thus neglect this effect in our simulations. The contribution of the SM states, which were neglected in Eq.~\ref{eq:1pi_im}, were taken into account in the numerical collider simulations.

\section{LHC constraints}\label{current_bound}

\begin{figure}[t]
	\centering
	\includegraphics[trim={6cm 20cm 8cm 3.9cm},clip,scale=0.6]{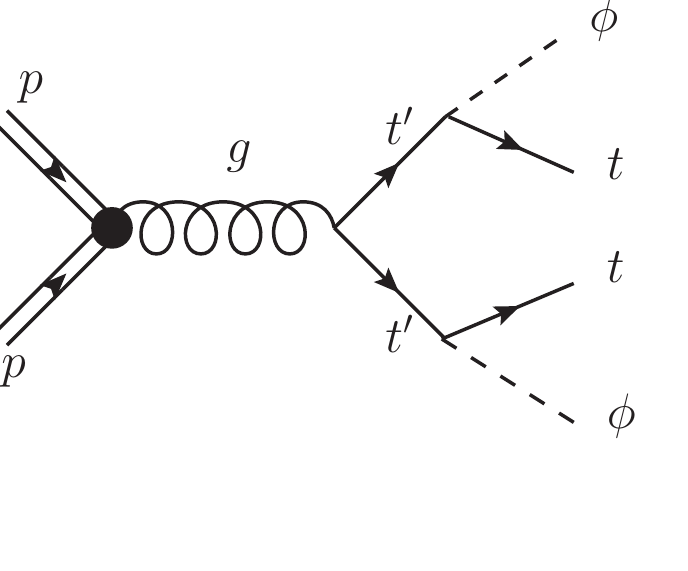}
	\caption{\small\it Representative Feynman diagrams for pair production of $t^{\prime}$ and its subsequent major decay.}
	\label{fig:feynman}
\end{figure}

\begin{figure}
	\centering
\subfloat[\label{fig:current_bound_gamma_mps100}]{\includegraphics[scale=0.4]{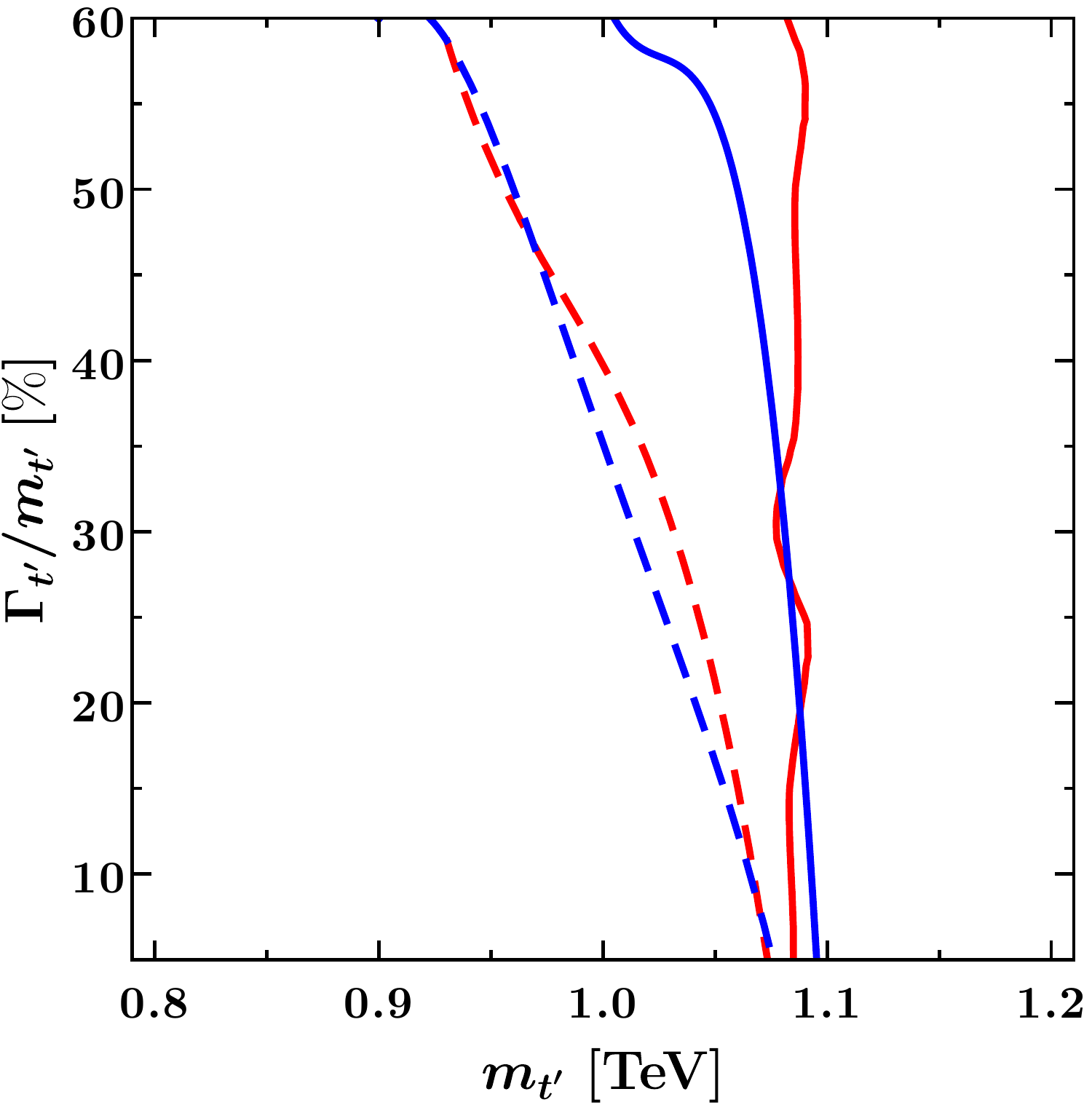}}
\hspace{1cm}
\subfloat[\label{fig:current_bound_gamma_mps300}]{
\includegraphics[scale=0.4]{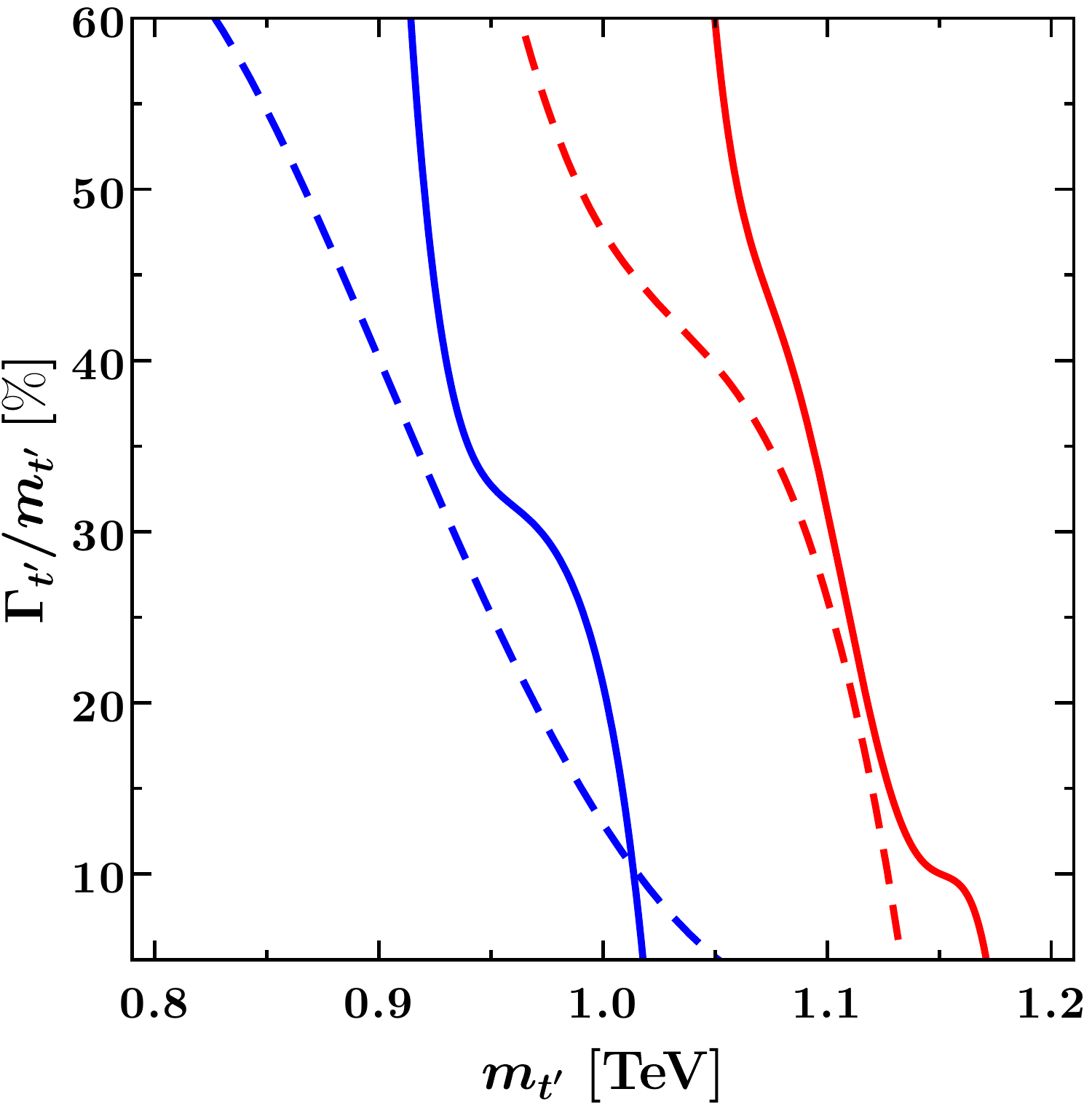}}
	\caption{\small\it Current bound on the $m_{t^{\prime}}-\Gamma_{t^{\prime}}/m_{t^{\prime}}$ parameter space from ATLAS $13$ TeV stop search \cite{ATLAS-CONF-2017-019} (blue) and the ATLAS $13$ TeV VLQ search \cite{Aaboud:2018xuw} (red) at $36.1~fb^{-1}$ integrated luminosity. Exclusion contours for both 1PI (solid) and NWA (dashed) cases have been shown. $M_{\phi}$ has been fixed at (a) $100$ and (b) $300$ GeV.}
	\label{fig:current_bound}
\end{figure}

In this section we recast the current null results from LHC 13 TeV runs on the parameter space of the toplike vector quark $t^{\prime}$ introduced in the previous section. In the parameter space of interest the branching ratio $Br(t^{\prime}\rightarrow\phi t)$ always remains greater than $80\%$. The dominant pair production process of $t^{\prime}$ are depicted in Fig.~\ref{fig:feynman}. For $M_{\phi}<2m_t$, $\phi$ dominantly decays to a pair of $b$ quarks ($Br(\phi\rightarrow b\bar{b}>90\%)$). Once the $t\bar{t}$ decay mode becomes kinematically accessible the branching ratio to a pair of tops starts to dominate. In our analysis we set the benchmark values of the pseudoscalar mass below the top pair threshold. Hence the most optimistic final state of interest includes a top pair and at least two $b$ jets. We concentrate on possible $13$ TeV results from CMS and ATLAS that can constraint this final state to put bounds on the parameter space of the exotic states.\\\\
To numerically study the current results from the LHC studies we implement the effective Lagrangian defined in Eqs. \ref{master:lgn} and \ref{master:lgn:a} in {\tt FeynRules 2.0} \cite{Alloul:2013bka}. We incorporate Eq.~\ref{eq:1pi_im} into {\tt MadGraph5} \cite{Alwall:2014hca} to obtain the 1PI corrected propagator for $t^{\prime}$ in our analysis. We generate leading order (LO) events for the process depicted in Fig.~\ref{fig:feynman} for both 1PI and BW scenarios in {\tt MadGraph5} not assuming $t^{\prime}$ to be on shell to incorporate the effect of the $t^{\prime}$ propagator. Higher order effects have been captured using a next to leading order (NLO) K-factor of $1.3$. The K-factor was obtained using the central value of top pair production cross section from {\tt Top++} \cite{Czakon:2011xx} by scaling the top mass to the top partner mass since they have identical SM quantum numbers and are expected to get similar QCD corrections when the masses are appropriately scaled. The obtained K-factor can vary up to $5\%$ with changes in factorization and renormalization scales. We have checked that the K-factor obtained from {\tt Top++} is within $10\%$ of the value extracted from {\tt MadGraph} at NLO. For our intended level of accuracy a cross section scaling with a K-factor provides a fairly good approximation of the bounds at NLO since the kinematic shapes and hence the efficiencies have a soft dependence on higher order effects as shown in Appendix~\ref{appn:nlo}. We parton shower the events using {\tt Pythia8} \cite{Sjostrand:2014zea} and jet-cluster with {\tt FastJet} \cite{Cacciari:2011ma} using the anti-$k_T$ algorithm \cite{Cacciari:2008gp}. Detector efficiencies are incorporated using the default ATLAS card of {\tt Delphes} \cite{deFavereau:2013fsa}.\\\\
We find the latest bounds on the $m_{t^{\prime}}-\Gamma_{t^{\prime}}/m_{t^{\prime}}$ parameter space from all $13$ TeV CMS and ATLAS analysis implemented in {\tt CheckMate 2.0} \cite{Dercks:2016npn} and from the single lepton channel VLQ searches from \cite{Aaboud:2018xuw} recasted in {\tt MadAnalysis5} \cite{Conte:2012fm}. The validation of our recast of the single lepton studies of the VLQ search \cite{Aaboud:2018xuw} is summarized in Appendix~\ref{appn:validation}. We vary $m_{t^{\prime}}$ and $g_{\phi}^{\ast}$ for fixed values of $M_{\phi}$ (see Table~\ref{tab:parameters}). For each point in the parameter space the intrinsic {\tt Checkmate 2.0} parameter $R$\footnote{The R parameter is defined as $R\equiv(S-1.64\Delta S)/S95$ where $S$ is the expected number of signal events in a particular signal region with uncertainty $\Delta S$ and $S95$ is the allowed number of signal events at $95\%$ confidence level in that signal region.} is calculated using all implemented LHC 13 TeV analyses. The $R=1$ contour provides the $2\sigma$ exclusion on our parameter space. To obtain the constraints from the ATLAS VLQ search \cite{Aaboud:2018xuw} we find the signal efficiencies $\epsilon$ defined as the fraction of the initial number of generated events that survive after applying the cuts defined in the ATLAS analysis for each point in the parameter space. We obtain the expected number of events at $\mathcal{L}=36.1~fb^{-1}$
\begin{equation}
N=\sigma^{LO}\times K\times\epsilon\times\mathcal{L}
\label{eq:nsim}
\end{equation}
where $\sigma^{LO}$ is LO cross section obtained from {\tt MadGraph5} and $K$ is the NLO K factor. The exclusion obtained from \cite{Aaboud:2018xuw} by requiring the expected number of signal events to be within the SM uncertainty at $2\sigma$ and the {\tt Checkmate} exclusion in the $m_{t^{\prime}}-\Gamma_{t^{\prime}}/m_{t^{\prime}}$ parameter space are plotted in Fig.~\ref{fig:current_bound}. Essentially we vary $g_{\phi}^{\ast}$ to tune the decay width $\Gamma_{t^{\prime}}/m_{t^{\prime}}$. Increase in $g_{\phi}^{\ast}$ has two effects on the cross section, an enhancement because of increase in the $t^{\prime}\rightarrow\phi t$ branching ratio and a propagator suppression due to increase in the width of $t^{\prime}$. The change in constraints on $m_{t^{\prime}}$ as seen in Fig.~\ref{fig:current_bound} is a result of the interplay between these two conflicting effects.
\begin{table} 
	\centering
	\begin{tabular}{ cccc }
		\hline\hline
		$m_{t^{\prime}}$ & $M_{\phi}$ & $g_{\phi}^{\ast}$ & $\sin\theta_R$ \\ 
		\hline
		$0.8-1.3$ TeV & $100,300$ GeV & $0.1-10$ & $0.1$ \\ 
		\hline\hline
	\end{tabular}
	\caption{\small\it Choice of free parameters}
	\label{tab:parameters}
\end{table}
As can be seen from the plot, with increasing $\Gamma_{t^{\prime}}/m_{t^{\prime}}$, the 1PI contour deviates from the NWA one and is more constraining. The {\tt Checkmate} analysis which provides the most aggressive constraint on our parameter space is an ATLAS stop search at $36.1~fb^{-1}$ integrated luminosity that searches for a channel consisting of at least four $b$-jets, $1-2$ leptons and $\slash{E}_T$ \cite{ATLAS-CONF-2017-019}. This is very similar to the final channel we expect when the final state tops decay leptonically. For the VLQ searches \cite{Aaboud:2018xuw} the signal region with at least two top quarks and at least four $b$-jets is of interest. The constraints from \cite{Aaboud:2018xuw} is stricter since it focuses on the hadronic decays of the top quark which has more branching ratio ($>60\%$). For $M_{\phi}=100$ GeV, the pseudoscalar $\phi$ mimics a Higgs which is a focus of the  ATLAS stop search \cite{ATLAS-CONF-2017-019}. Thus the constraints from this search is reduced when we take $M_{\phi}=300$ GeV as can be seen from Fig.~\ref{fig:current_bound_gamma_mps300}.

\section{HL-LHC Reach}\label{sec:future_strategy}

In this section we present an optimized search strategy for the vectorlike quark $t^{\prime}$ at the HL-LHC. The HL-LHC is expected to run at a centre of mass energy of $14$ TeV and collect data till $3~ab^{-1}$ of integrated luminosity \cite{Cepeda:2019klc}. We focus on the $pp\rightarrow t^{\prime}\bar{t}^{\prime}\rightarrow\phi\phi^{\ast}t\bar{t}$ signal topology where the pseudoscalar $\phi$ predominantly decays to a pair of bottom quarks. The relevant SM backgrounds are $t\bar{t}b\bar{b}$, $t\bar{t}+Z$ and $t\bar{t}+H$. The NLO K-factors for these channels are obtained from \cite{Alwall:2014hca} and reported after multiplying with the LO cross sections obtained from {\tt MadGraph5} in Table~\ref{tab:efficiencies}.

\subsection{Cut based analysis}

\begin{figure}[t]
	\centering
	\subfloat[\label{fig:met}]{\includegraphics[scale=0.33]{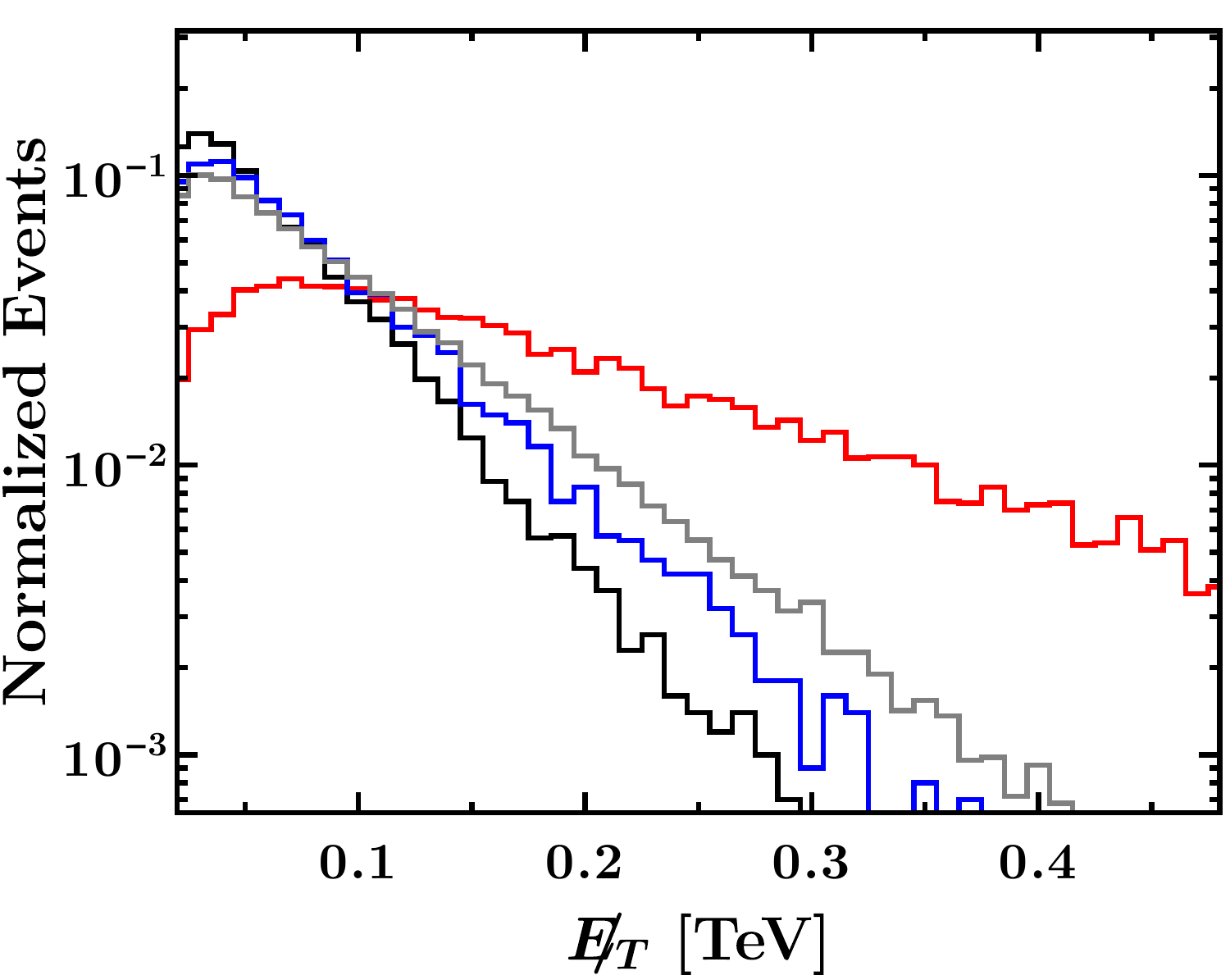}}
	\hspace{0.1cm}
	\subfloat[\label{fig:ptl1}]{\includegraphics[scale=0.33]{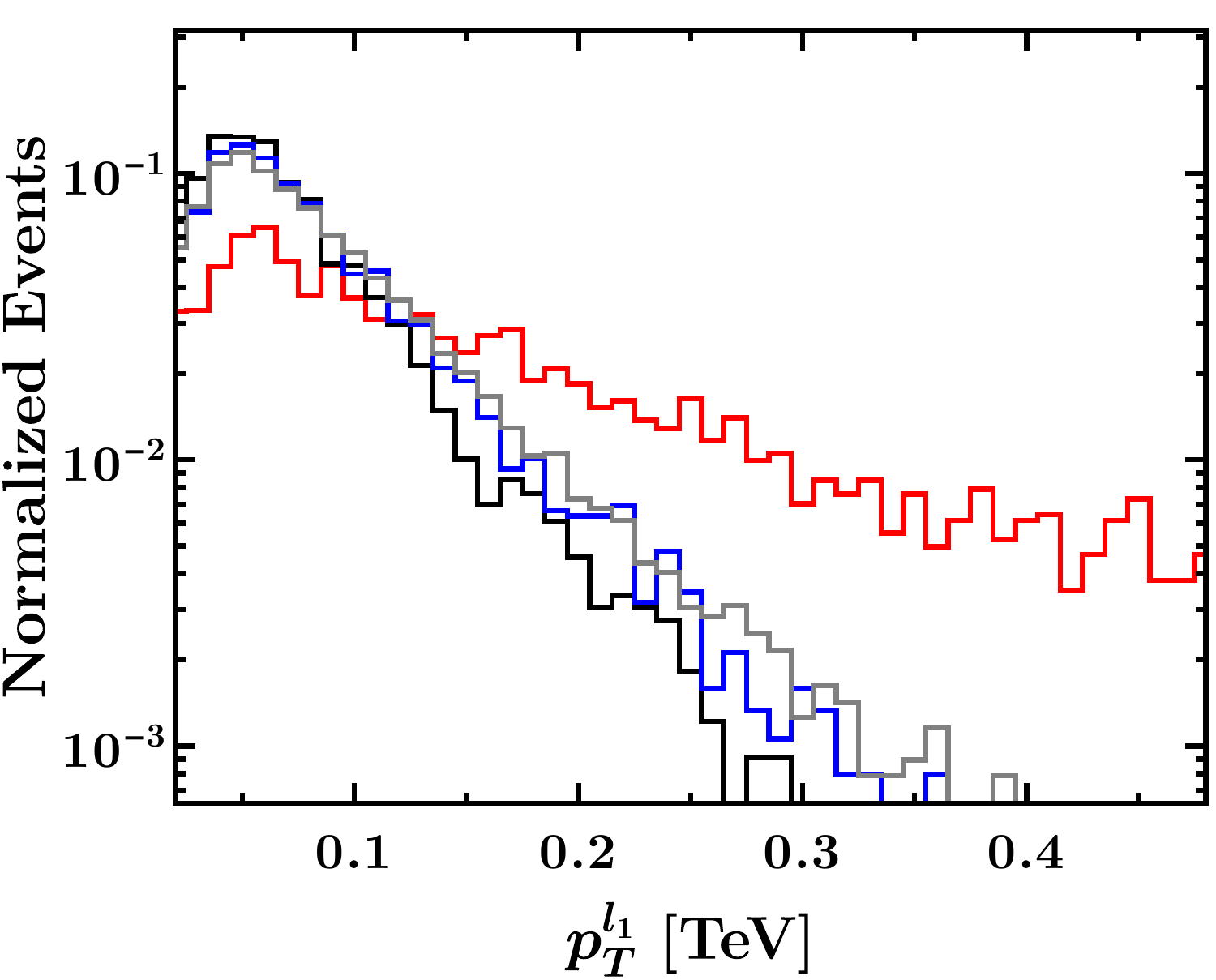}}
	\hspace{0.1cm}
	\subfloat[\label{fig:ptj1}]{\includegraphics[scale=0.33]{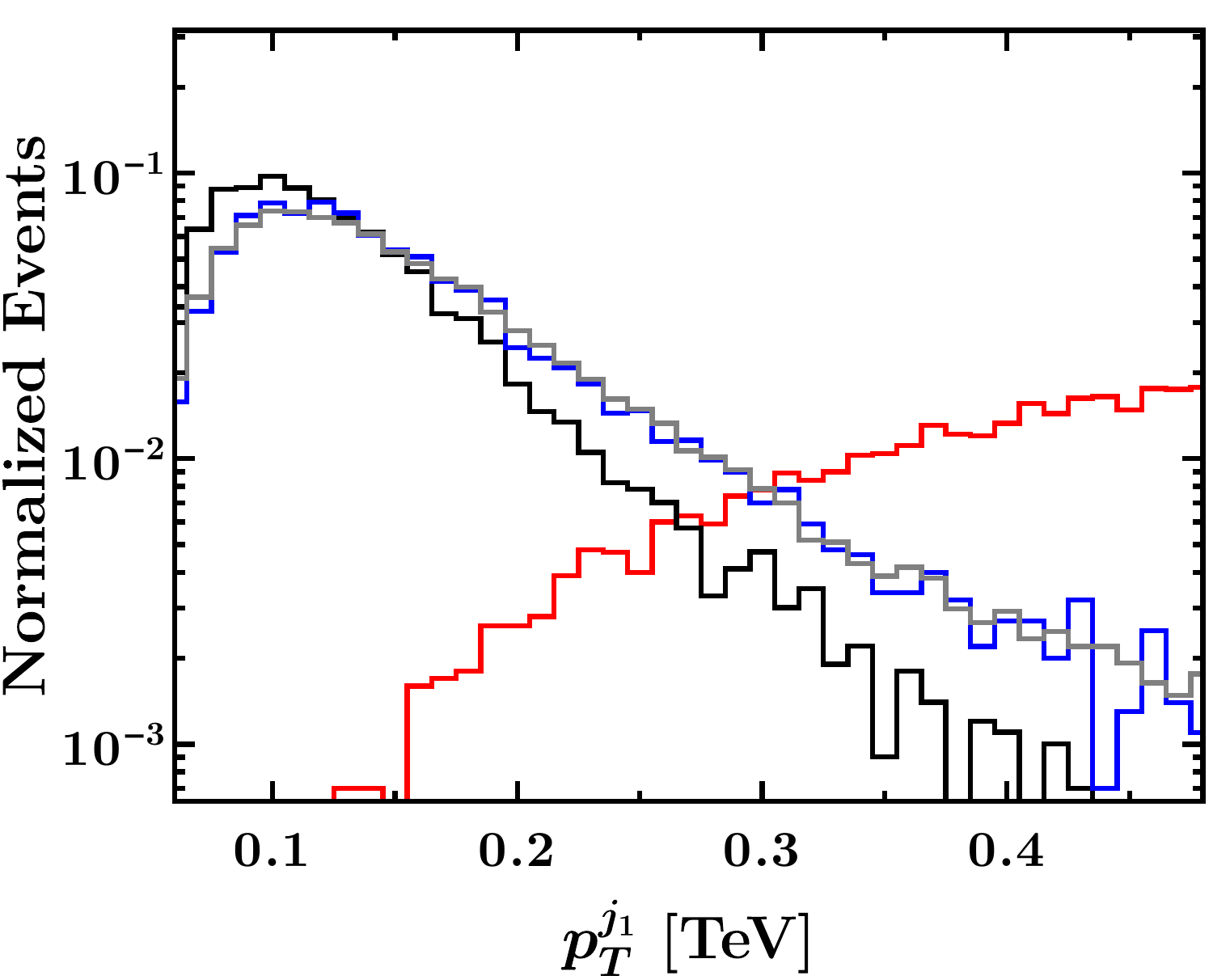}}\\
	\subfloat[\label{fig:ptb1}]{\includegraphics[scale=0.33]{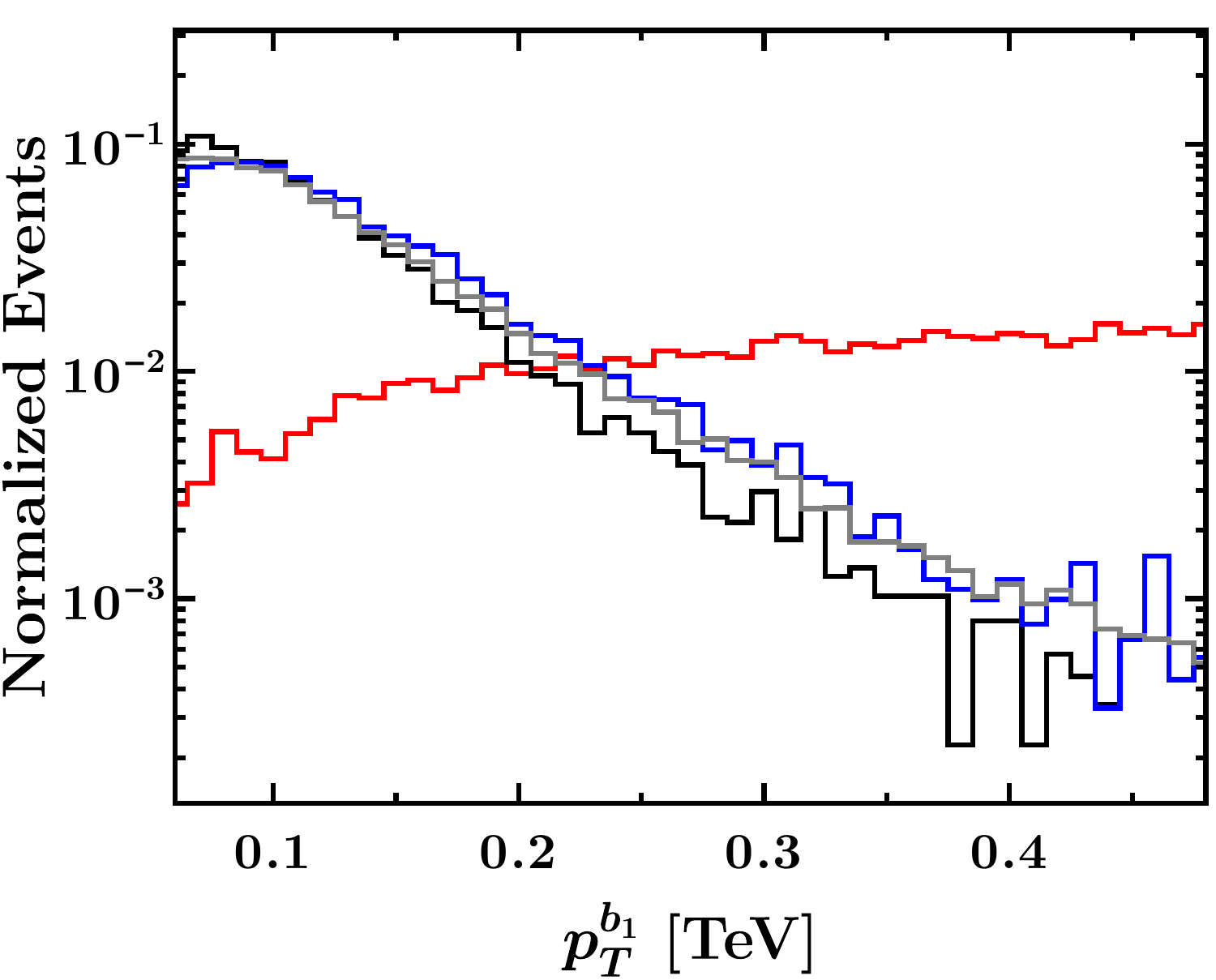}}
	\hspace{0.1cm}
	\subfloat[\label{fig:nj}]{\includegraphics[scale=0.33]{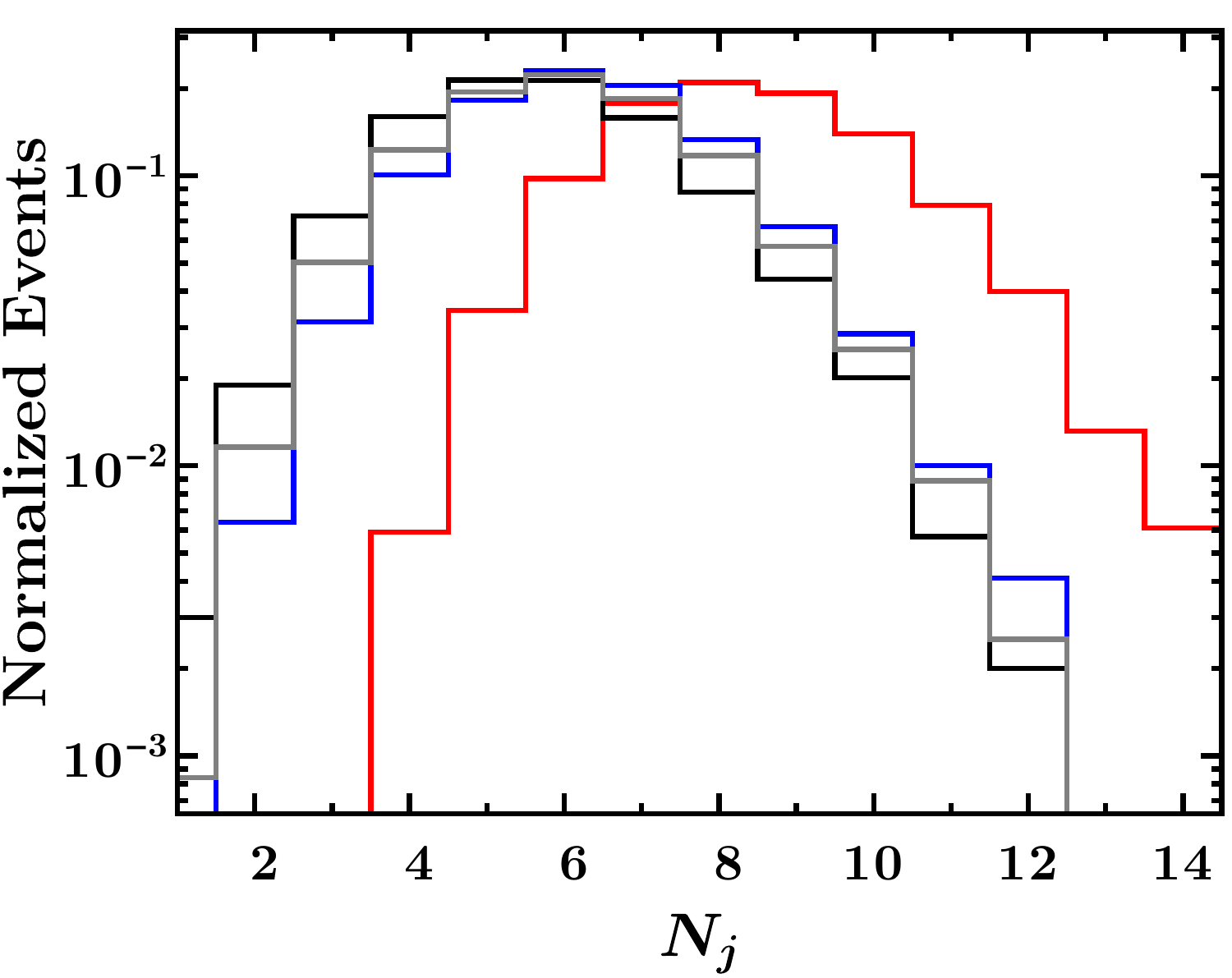}}
	\hspace{0.1cm}
	\subfloat[\label{fig:nb}]{\includegraphics[scale=0.33]{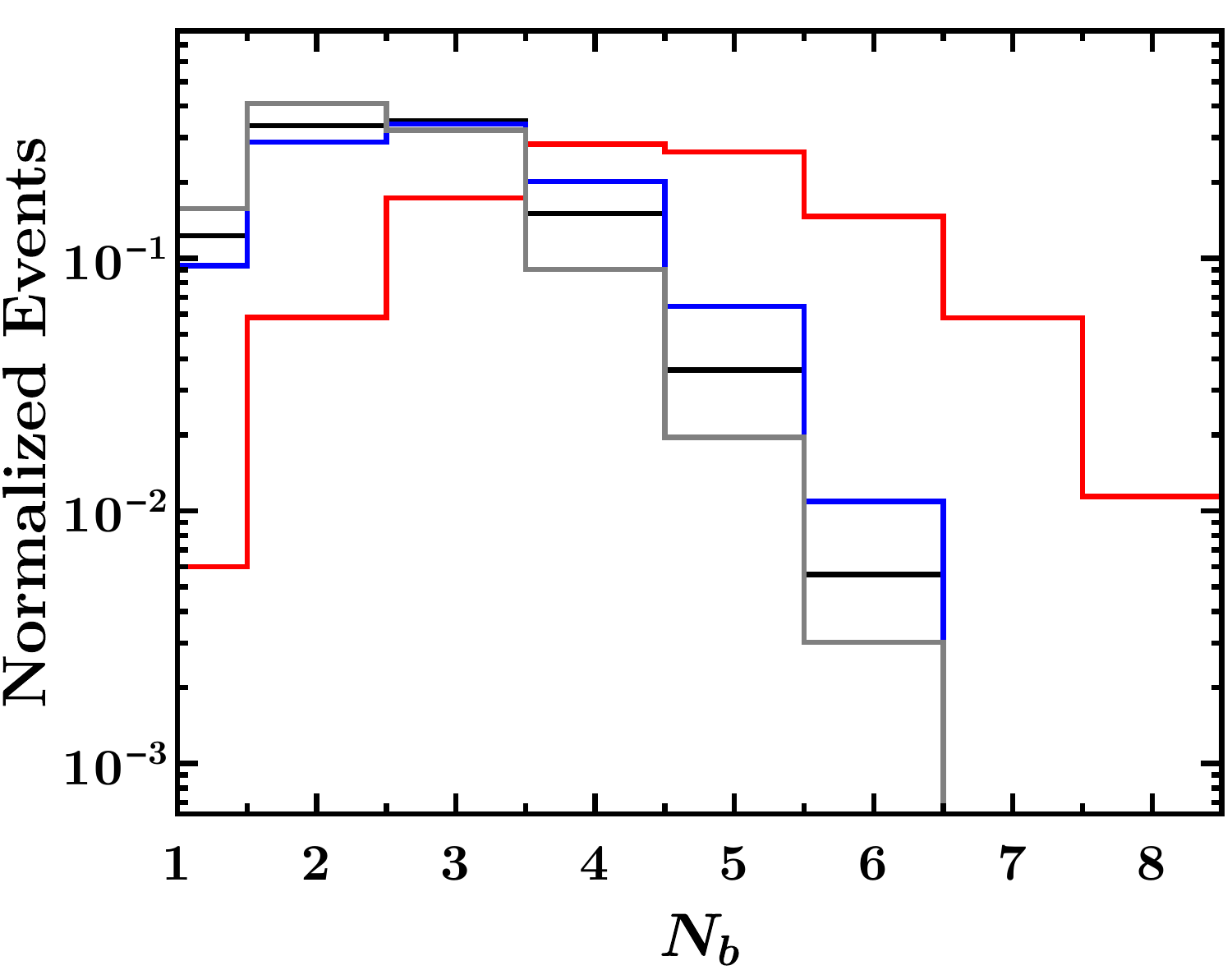}}\\
	\subfloat[\label{fig:mb1b2}]{\includegraphics[scale=0.33]{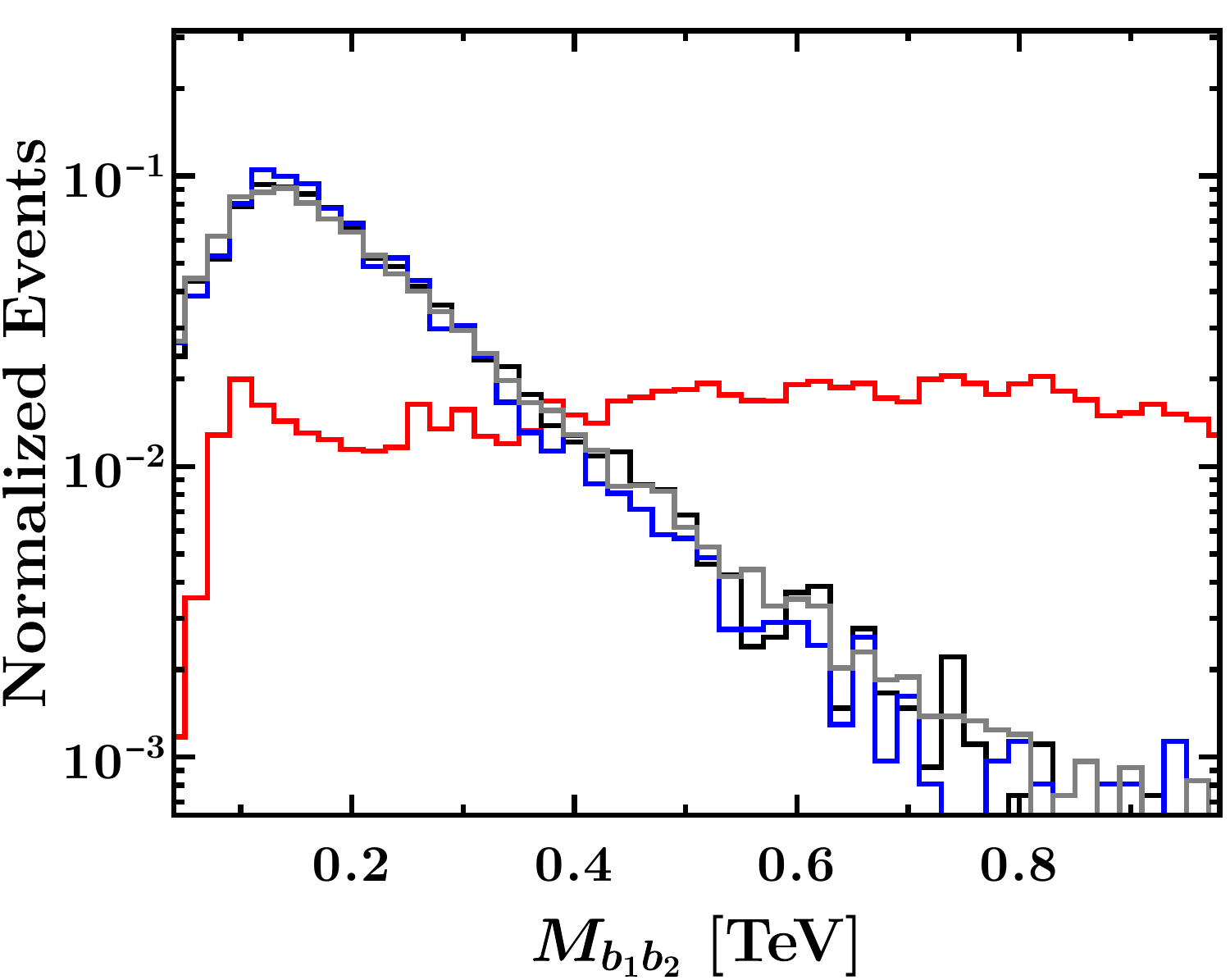}}
	\caption{\small\it Reconstructed level normalized event distributions for observables displaying significant enhancement for the 1PI signal (red) over the backgrounds $t\bar{t}b\bar{b}$ (black), $t\bar{t}+Z$ (grey) and $t\bar{t}+H$ (blue). For the 1PI signal we set $m_{t^\prime}$ at $1.1$ TeV, $M_{\phi}$ at $100$ GeV and $\Gamma_{t^{\prime}}/m_{t^{\prime}}$ at $0.5$.}
	\label{fig:distributions}
\end{figure}

To extract optimized cuts for the signal topology of interest a systematic study of the kinematic distributions for both the signal and backgrounds is now in order. In Fig~\ref{fig:distributions} we plot the signal and background distributions for various kinematic observables. All the distributions are normalized to one event. For the signal a benchmark point ($m_{t^\prime}=1.1$ TeV, $M_{\phi}=100$ GeV and $\Gamma_{t^{\prime}}/m_{t^{\prime}}=0.5$) that is not excluded by the current LHC bounds is chosen.
Taking cue from the plots in Fig.~\ref{fig:distributions} we device the reconstructed level cuts listed in Table~\ref{tab:recocuts}.
\begin{table}[t!]
	\centering
	\begin{tabular}{cccccccc}
		\hline\hline
		Observable & $\slashed{E}_{T,miss}$ & $p_T^{l_1}$ & $p_T^{b_1}$ & $p_T^{j_1}$ & $N_j$ & $N_b$ & $M_{b_1b_2}$  \\
		\hline
		Cut & $>100$ GeV & $>100$ GeV & $>200$ GeV & $>250$ GeV & $>6$ & $>3$ & $>350$ GeV \\
		\hline\hline 
	\end{tabular}
	\caption{\small\it Reconstructed level cuts used to distinguish the signal from backgrounds.}
	\label{tab:recocuts}
\end{table}
In order to obtain the projected reach of the HL-LHC with the optimized cuts we scale the parameter space between $m_{t^{\prime}}\in[0.8,1.8]$ TeV and $\Gamma_{t^{\prime}}/m_{t^{\prime}}\in[2.5\%,60\%]$. For each point in this parameter space we generate $2.5\times10^4$ signal events using the full 1PI propagator at $14$ TeV. We also similarly prepare $10^5$ background events for each of the backgrounds. We find out the signal and background efficiencies ($\epsilon_{S/B}$) for both our designed cuts and the ATLAS analysis \cite{Aaboud:2018xuw} that provides the strictest present bounds. We plot the $5\sigma$ contours at $3~ab^{-1}$ in Fig.~\ref{fig:reach}. The optimized cuts lead to $\sim5\%$ improvement in the reach of HL-LHC on the parameter space compared to the ATLAS search \cite{Aaboud:2018xuw}.

\subsection{Machine learning approach}

\begin{figure}[t]
	\centering
	\subfloat[\label{fig:roc}]{\includegraphics[scale=0.4]{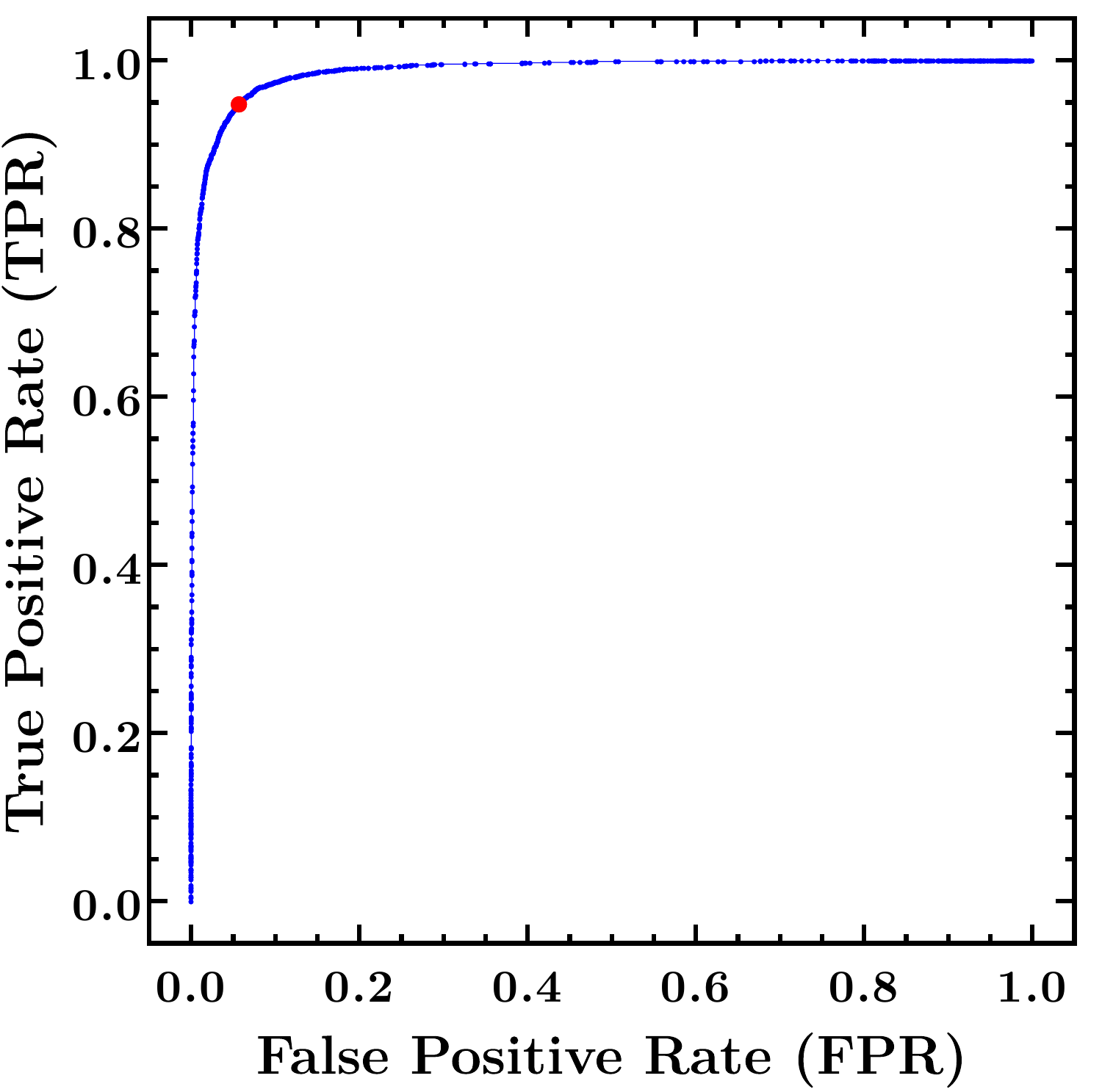}}
	\hspace{1cm}
	\subfloat[\label{fig:reach}]{\includegraphics[scale=0.4]{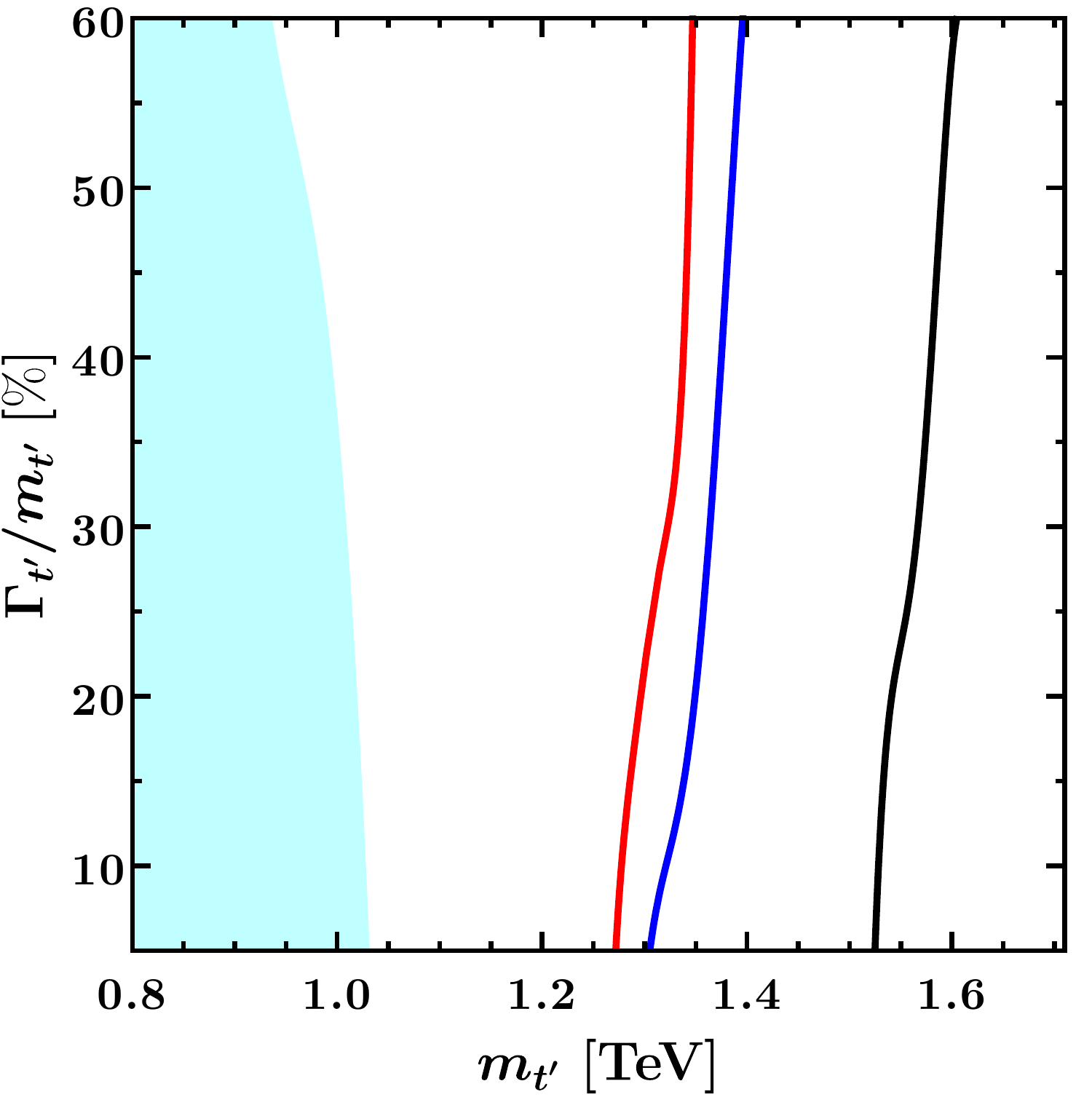}}
	\caption{\small\it (a) The ROC curve for $m_{t^\prime}=1.4$ TeV and $\Gamma_{t^{\prime}}/m_{t^{\prime}}=0.5$. The red point has the maximum difference between TPR and FPR. (b) The $5\sigma$ HL-LHC reaches from XGBoost (black,solid), ATLAS analysis \cite{Aaboud:2018xuw} (red,solid) and our designed cuts listed in Table~\ref{tab:recocuts} (blue,solid) for pair production of $t^{\prime}$. The blue region denotes the $13$ TeV excluded region from the ATLAS search \cite{Aaboud:2018xuw}. All reaches were estimated at $3~ab^{-1}$ integrated luminosity. We set $M_{\phi}$ at $100$ GeV.}
\end{figure}

Next we investigate the possibility to enhance the HL-LHC reach by utilizing ML techniques. We focus on the XGBoost algorithm to improve the signal to background discrimination in the parameter space of interest.\\\\
The XGBoost \cite{2016} algorithm is a framework available in multiple coding languages that can act as a classifier and a regressor. We use it as a binary classifier that can classify between two classes dependant on multiple features. The classifier model is trained using example datasets to optimize the cuts on these features to best distinguish between the classes. We have tuned various hyperparameters to optimize the classifier model with a training loss of $\sim 5\%$ and validation loss of $\sim6\%$, keeping overtraining in control.  \\\\
We use XGBoost to distinguish between the signal and background events using twenty features of the final state topology listed below.
\begin{itemize}
	\item The transverse momenta $p_T$ of : the two hardest leptons, four hardest jets and the four hardest b-jets.
	\item The invariant masses of all possible b-jet pairs from the four hardest b-jets.
	\item The missing transverse energy $\slash{E}_T$.
\end{itemize}
We prepare $2.5\times10^4$ background events for the three leading channels weighted according to their LO cross sections. For each point in the $m_{t^{\prime}}-\Gamma_{t^{\prime}}/m_{t^{\prime}}$ parameter space, we randomize the signal and background events to minimize the possibility of biasness and to increase the chance of mimicking a representative scenario which was fed as an input to XGBoost. We use $80\%$ of the data for training and the rest to test. The receiver operating characteristics (ROC) curve for a benchmark point in our parameter space ($m_{t^\prime}=1.4$ TeV, $M_{\phi}=100$ GeV and $\Gamma_{t^{\prime}}/m_{t^{\prime}}=0.5$) is shown in Fig.~\ref{fig:roc}. Here the true positive rate (TPR) signifies the signal efficiencies and the false positive rate (FPR) signifies the background efficiencies. For each point in the parameter space we identify the TPR and FPR from the ROC curves with the maximum difference and their corresponding values represent the selected signal and background efficiencies ($\epsilon_{S/B}$) respectively for that point in the parameter space. We plot the $5\sigma$ HL-LHC reach using the ML efficiencies in Fig.~\ref{fig:reach} along with the corresponding reaches from the ATLAS VLQ search \cite{Aaboud:2018xuw} and the optimized cut analysis. As can be seen from Fig.~\ref{fig:reach} ML techniques significantly improves the HL-LHC reach ($\sim19\%$) compared to the ATLAS search \cite{Aaboud:2018xuw} reaching up to $1.6$ TeV for $\Gamma_{t^{\prime}}/m_{t^{\prime}}=0.6$ by optimizing cuts on all relevant kinematic observables. Details of the efficiencies and cross sections for the benchmark point ($m_{t^\prime}=1.1$ TeV, $M_{\phi}=100$ GeV and $\Gamma_{t^{\prime}}/m_{t^{\prime}}=0.5$) and the backgrounds is given in Table~\ref{tab:efficiencies}. The reach from single production channels are highly suppressed due to their low cross section as shown in Appendix~\ref{appn:single_reach}.
\begin{table}
	\centering
	\begin{tabular}{ccccc}
		\hline\hline
		\multirow{2}{*}{Channel} & \multirow{2}{*}{Cross section $\times$ K-factor (pb)} & \multicolumn{3}{c}{Signal Efficiency}\\
		& & ATLAS VLQ search & Table~\ref{tab:recocuts} Cuts & XGBoost\\
		\hline
		1PI signal & $0.029$ & $0.0163$ & $0.0137$ & $0.98$ \\
		$t\bar{t}b\bar{b}$ & $39.46$ & $4\times10^{-5}$ & $1\times10^{-5}$ & \multirow{3}{*}{$0.03$} \\
		$t\bar{t}+Z$ & $1.02$ & $6\times10^{-5}$ & $4\times10^{-5}$ & \\
		$t\bar{t}+H$ & $0.515$ & $1\times10^{-4}$ & $1\times10^{-4}$ & \\
		\hline\hline
	\end{tabular}
	\caption{\small\it $14$ TeV LO cross sections times NLO K-factors and the efficiencies from the ATLAS VLQ search \cite{Aaboud:2018xuw}, our new cut-design and from XGBoost for the 1PI signal and the backgrounds $t\bar{t}b\bar{b}$, $t\bar{t}+Z$ and $t\bar{t}+H$. For the 1PI signal we set $m_{t^\prime}$ at $1.1$ TeV, $M_{\phi}$ at $100$ GeV and $\Gamma_{t^{\prime}}/m_{t^{\prime}}$ at $0.5$.}
	\label{tab:efficiencies}
\end{table}

\section{Analyzing broad resonances at Colliders}\label{sec:prediction}

\begin{figure}[t]
	\centering
	\subfloat[\label{fig:resonance}]{\includegraphics[scale=0.4]{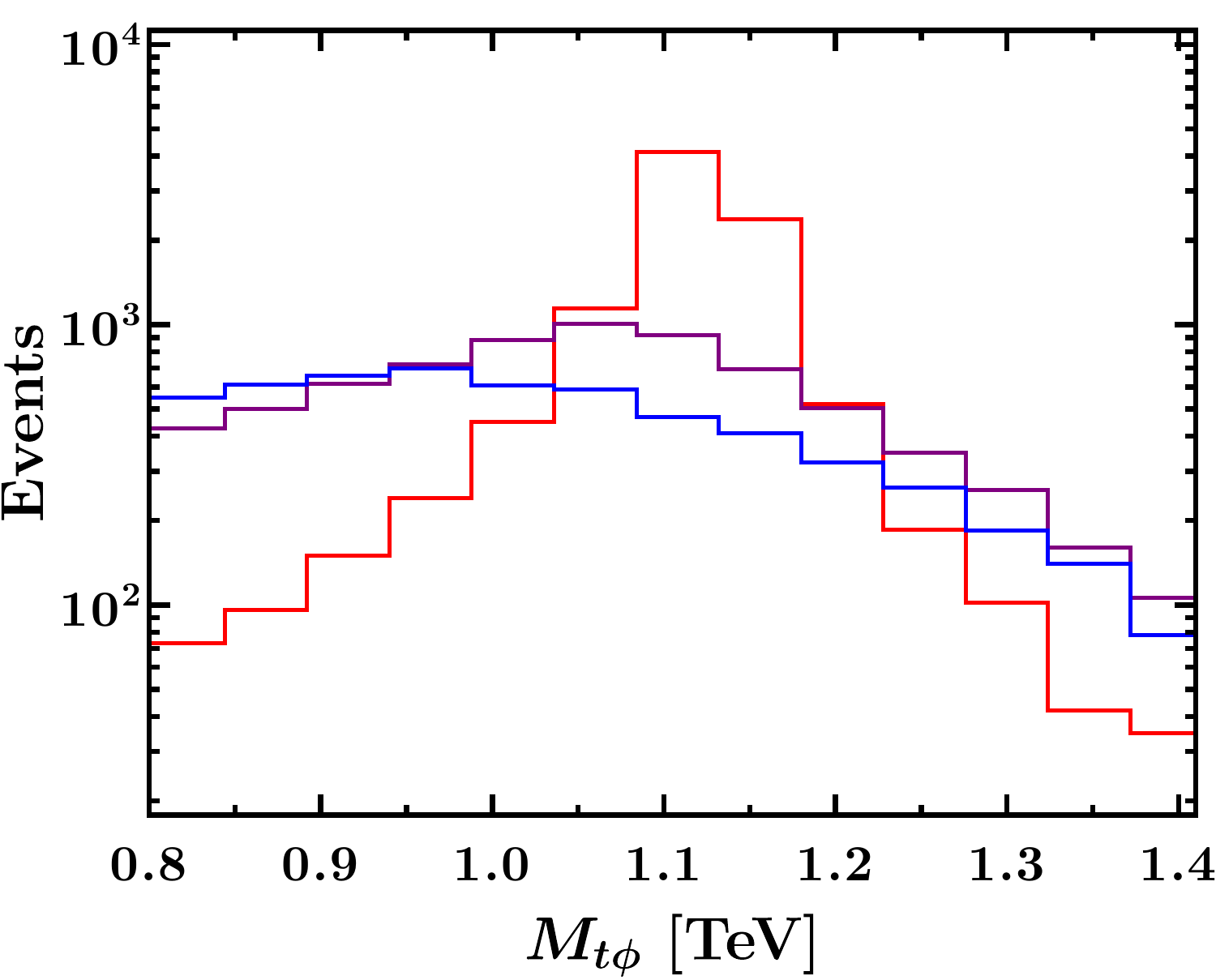}}\hspace{1cm}
	\subfloat[\label{fig:mass_pred_error}]{\includegraphics[scale=0.35]{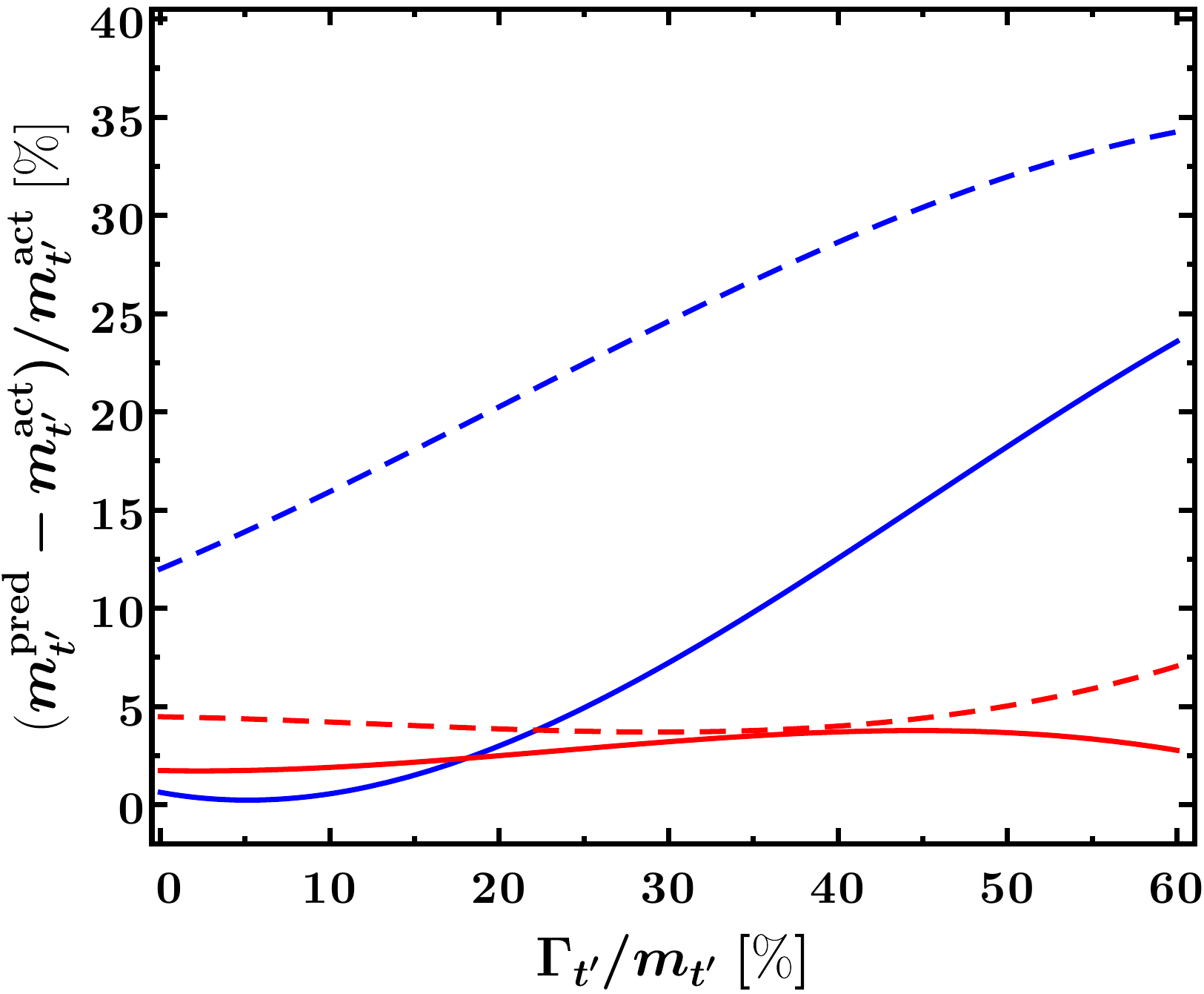}}
	\caption{\small\it (a) Invariant mass of pseudoscalar $\phi$ and top ($M_{t\phi}$) decaying from a top partner $t^{\prime}$ for $m_{t^{\prime}}=1.1$ TeV, $M_\phi=0.1$ TeV and $\Gamma_{t^{\prime}}/m_{t^{\prime}}=5\%$ (red), $30\%$ (purple) and $50\%$ (blue). (b) Relative deviation of the predicted mass $m_{t^{\prime}}^{pred}$ from the actual mass $m_{t^{\prime}}^{act}$ using BW fit (blue) and XGBoost (red). The solid line is for fitting done in the parton level and the dashed line is for fitting done in the reconstructed level. }
	\label{fig:ML_mass_pred}
\end{figure}

We now turn our attention to the question of extracting physical parameters, like the mass of the toplike top partner, from a broad resonance peak showing up at collider searches in the future. Expectedly the resonance peak gets deformed as the width to mass ratio increases and extracting the top partner mass from the resonance may be nontrivial. This can be seen from Fig.~\ref{fig:resonance} where we plot the invariant mass of the top and pseudoscalar $\phi$ decaying from the top partner after $t^{\prime}$ is pair produced through the process shown in Fig.~\ref{fig:feynman}. These are parton level results used for clear demonstrations of the effects of broadness on the resonance peaks. As can be seen from the plot, with increase in $\Gamma_{t^{\prime}}/m_{t^{\prime}}$ the peak shifts toward a lower value of top partner invariant mass \cite{Liu:2019bua} thus making the traditional approach using BW fitting prone to large errors. In this scenario ML algorithms may provide an useful handle to address these issues. As a proof of principle we employ the ML techniques on the arguably more challenging pair production channel to extract the mass of the top partner. We comment on the relative efficacy of this approach over traditional analysis methods of such resonance peaks. In principle a similar approach also works for single production of $t^{\prime}$.
We scan the parameter space between $m_{t^{\prime}}\in[0.6,1.8]$ TeV and $\Gamma_{t^{\prime}}/m_{t^{\prime}}\in[2.5\%,60\%]$ and for each point in this parameter space we generate $t^{\prime}$ pair production events as shown in Fig.~\ref{fig:feynman}. To check our accuracy in parton level we generate histograms by reconstructing the vectorlike quark mass by taking the invariant mass of top and the pseudoscalar $\phi$ decaying from $t^{\prime}$. For detector level results we allow one of the top quarks to decay hadronically and the other to decay leptonically. We use the detector level events to reconstruct the vectorlike quark mass as described below and plot its histogram.
\begin{enumerate}
	\item Jets originating from bottom quarks can be $b$-tagged at the collider. We name the pair of non-$b$-tagged jets with invariant mass closest to the $W$ boson mass as jets that have decayed from the $W$ boson. We call them $jWtag1$ and $jWtag2$.
	\item We name the $b$ jet which along with $jWtag1$ and $jWtag2$ has invariant mass closest to the top mass as the $b$ jet decaying from the hadronically decaying top. We call it $b_{had}$.
	\item We find the longitudinal component of the neutrino momentum by requiring the signal lepton-$\slash{E}_T$ invariant mass to be equal to the $W$ boson mass.
	\item We name the $b$ jet which along with the signal lepton and the neutrino has invariant mass closest to the top mass as the $b$ jet decaying from the leptonically decaying top. We call it $b_{lep}$.
	\item We classify the four $b$ jets decaying from the two pseudoscalars by minimizing the mass difference from all possible $b$ jet pairs. We call the $b$ jets decaying from one $\phi$ as $b1tag1$ and $b2tag1$. We call the $b$ jets decaying from the other $\phi$ as $b1tag2$ and $b2tag2$.
	\item We identify the the $\phi$ that decayed along with the hadronically decaying top by minimizing the difference in invariant masses obtained by considering $b1tag1$ and $b2tag1$ along with $b_{had}$ and $jWtag1$ and $jWtag2$, and $b1tag2$ and $b2tag2$ along with $b_{lep}$ and the signal lepton and the neutrino, and vice versa. We call it $\phi_{had}$.
\end{enumerate}
We use the invariant mass of $\phi_{had}$, $b_{had}$, $jWtag1$ and $jWtag2$ to reconstruct the $t^{\prime}$ mass. The data from the histograms are used to train the XGBoost regressor. We use $80\%$ of the data to train the ML algorithm and the rest to test the accuracy of predicting the reconstructed mass of the broad resonance. We plot the prediction accuracy defined as the relative deviation of the predicted mass $m_{t^{\prime}}^{pred}$ from the actual mass $m_{t^{\prime}}^{act}$ for the traditional and ML techniques in Fig.~\ref{fig:mass_pred_error}. In Table~\ref{tab:mass_pred} we report the predicted top partner mass using XGBoost and the traditional method for different values of $\Gamma_{t^{\prime}}/m_{t^{\prime}}$. As can be seen from Fig.~\ref{fig:mass_pred_error}, the error in $m_{t^{\prime}}$ prediction using XGBoost stays within $5\%$ for any value of $\Gamma_{t^{\prime}}/m_{t^{\prime}}$ whereas the error using the traditional approach increases with the value of $\Gamma_{t^{\prime}}/m_{t^{\prime}}$ reaching up to $35\%$ ($25\%$) for $\Gamma_{t^{\prime}}/m_{t^{\prime}}=60\%$ in the reconstructed (parton) level. This clearly indicates that careful reconstruction of resonance peak using ML techniques like the one demonstrated here is imperative for the analysis of broad resonances. The prediction for the top partner width does not improve considerably when we use the ML technique. This is expected since the $t^{\prime}$ propagator and hence the shape of resonance peaks do not explicitly depend on the width as can be seen from Eq.~\ref{eq:1pi_im}.

\section{Conclusion}

In this paper we revisit the collider phenomenology of toplike vector quarks taking into account the effect of a large decay width. We consider a phenomenological model where the exotic vector quark preferentially decays to a pseudoscalar and a top quark. The large coupling between the toplike vector quark and the pseudoscalar drives the large decay width of the exotic fermion beyond the Breit-Wigner approximation. We use the 1PI corrected propagator to capture the leading effects of the broadness of the resonance. While this framework can readily arise from composite Higgs frameworks where the toplike vector quarks originate from the dynamics of a strong sector and is essential to generate radiative masses for the pNGB composite Higgs through partial compositeness \cite{Contino:2010rs} we mostly remain agnostic to the origin of the exotic states and the effective Lagrangian.\\\\
We use the studies implemented in {\tt Checkmate 2.0} and the ATLAS VLQ search \cite{Aaboud:2018xuw} to explore the present LHC constraints on the parameter space of this model. We find the present limits on the vector quark mass remains in the range of $0.9-1.15$ TeV for $\Gamma_{t^{\prime}}/m_{t^{\prime}}$ between $5\%$ and $60\%$ at $36.1~fb^{-1}$ integrated luminosity. Next we demonstrated that for future searches the ML technique that utilizes extreme gradient boosting is far more adept in increasing the reach of HL-LHC. With the ML technique we report a $19\%$ improvement in the discovery potential of vectorlike quark pair production over current cut based searches ranging between $1.5-1.6$ TeV for width to mass ratio in the range $5\%-60\%$ at $3~ab^{-1}$.\\\\
We briefly address the issue of extracting masses from distorted resonance peaks for a broad exotic vector quark. Predictably the inferences from traditional Breit-Wigner fitting starts getting erroneous as the width to mass ratio grows beyond the NWA. We show that optimized ML techniques may be better suited in this scenario.  
\begin{table}
	\centering
	\begin{tabular}{cccccc}
		\hline\hline
		\multirow{2}{*}{True Mass (TeV)} & \multirow{2}{*}{Predicting method} & \multirow{2}{*}{Analysis Level} & \multicolumn{3}{c}{$\Gamma_{t^{\prime}}/m_{t^{\prime}}$} \\
		\multirow{4}{*}{$0.115$}&&& $0.05$ & $0.3$ & $0.5$ \\
		\hline
		&\multirow{2}{*}{Traditional BW fit} & Parton Level & $0.115$ & $0.107$ & $0.919$ \\
		&& Reconstructed Level & $0.942$ & $0.862$ & $0.787$ \\
		&\multirow{2}{*}{XGBoost} & Parton Level & $0.113$ & $0.119$ & $0.119$ \\
		&& Reconstructed Level & $0.124$ & $0.118$ & $0.119$ \\
		\hline\hline
	\end{tabular}
	\caption{\small\it Predicted values of $m_{t^{\prime}}$ using ML and traditional BW fit methods for different values of $\Gamma_{t^{\prime}}/m_{t^{\prime}}$.}
	\label{tab:mass_pred}
\end{table}

\paragraph*{Acknowledgements\,:}
We thank Biplob Bhattacherjee, Atri Dey and Avik Banerjee for discussions. T.S.R. acknowledges Department of Science and Technology, Government of India, for support under Grant Agreement No. ECR/2018/002192 [Early Career Research Award]. S.D. and R.P. acknowledges MHRD, Govt. of India for the research fellowship. T.S.R. acknowledges the hospitality of ICTP, Trieste under the Associateship programme during the completion of this work.

\appendix

\section{NLO dependence of kinematic shapes}
\label{appn:nlo}

\begin{figure}[t]
	\centering
	\subfloat[\label{fig:nlo_met_ttz}]{\includegraphics[scale=0.4]{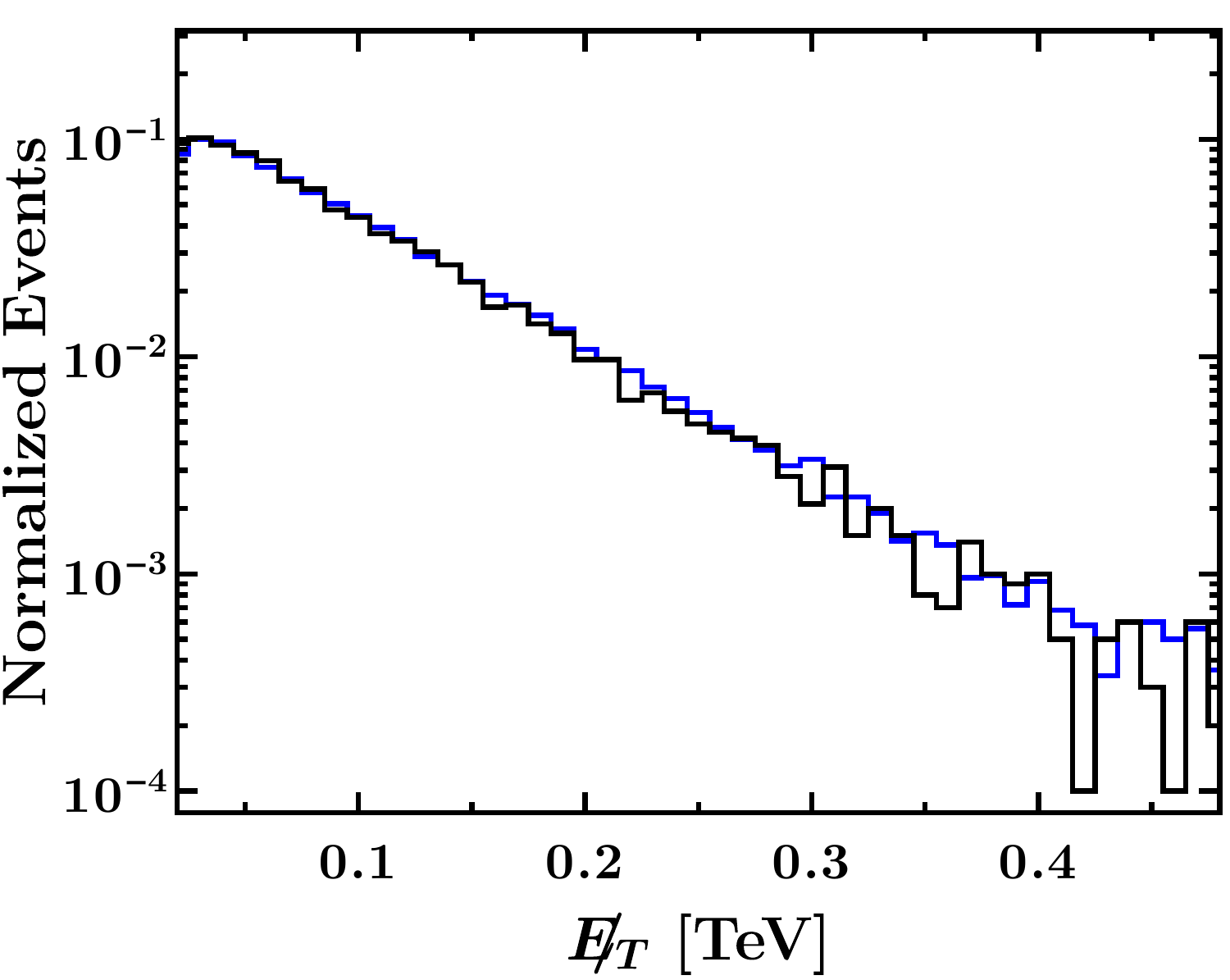}}
	\hspace{1cm}
	\subfloat[\label{fig:nlo_met_tt}]{\includegraphics[scale=0.4]{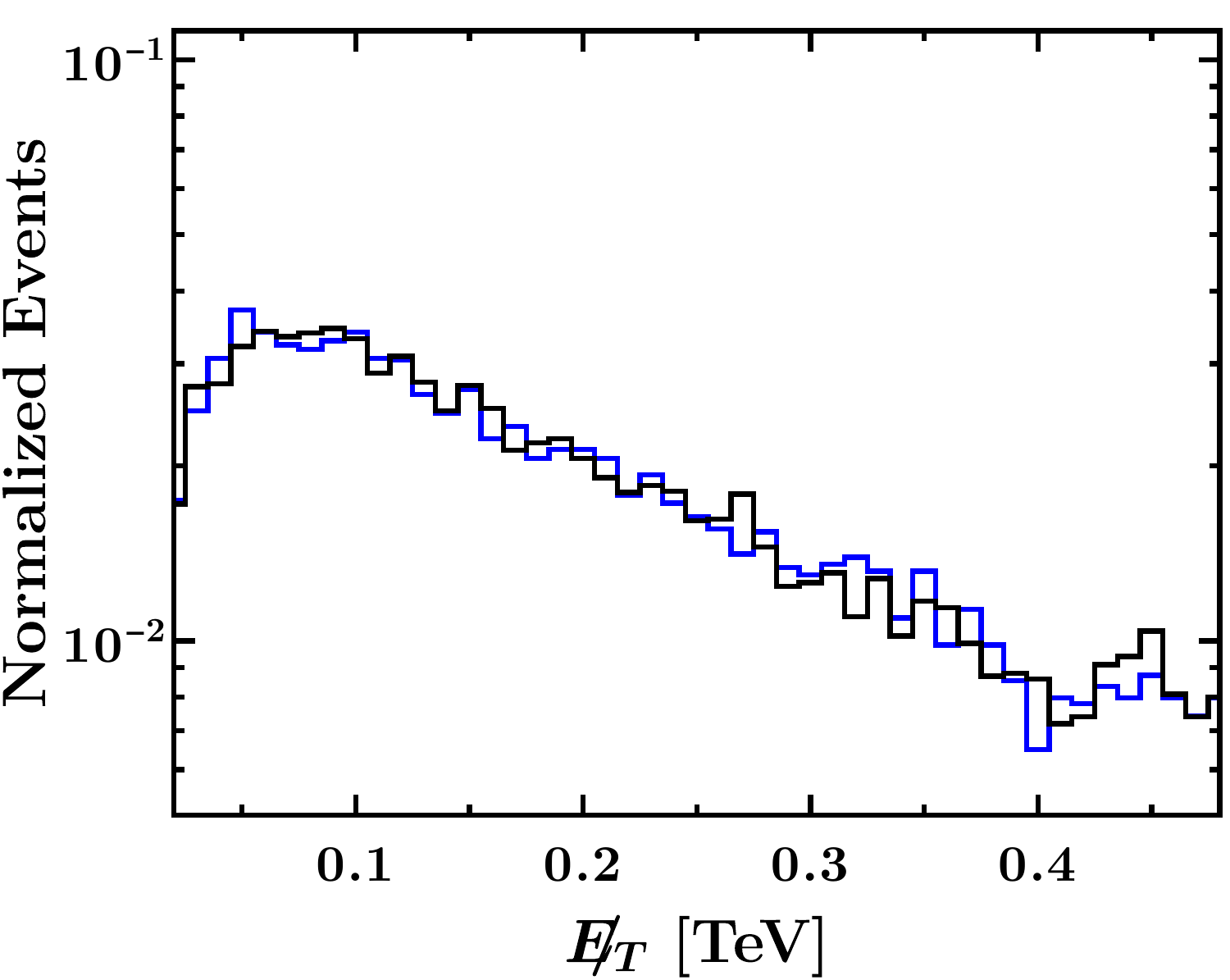}}
	\caption{\small\it Reconstructed level normalized event distributions for $t\bar{t}+Z$ (left) and $t\bar{t}$ for $m_t=1$ TeV (right) with events generated at LO (blue) and NLO using {\tt MadGraph} aMC@NLO (black). }
	\label{fig:nlo_distributions}
\end{figure}

In this section we demonstrate that the kinematic shapes depend mildly on higher order effects. We generated NLO background and signal events in {\tt MadGraph} aMC@NLO using the SM NLO UFO available along with the package. We simulated $t\bar{t}$ events by scaling the top mass to $1$ TeV to mimic the signal events used in the manuscript. Simulating the full signal will require the formulation and incorporation of higher order effective couplings in the {\tt MadGraph} UFO. Events were parton showered using {\tt Pythia8} and jet-clustered using {\tt FastJet}. Fast detector simulation was performed using {\tt Delphes} and the histograms were generated in {\tt MadAnalysis}. We have similarly produced the kinematic shapes for signal and background events at LO. In Fig.~\ref{fig:nlo_distributions} we show a couple of representative plots for both LO and NLO cases. As can be seen from the plot the NLO corrections in the region of interest is $\sim3\%$ which is lower than the expected systematic uncertainties and will have negligible effect on the cuts and efficiencies. A cross section scaling with a K-factor thus provides a good estimate of the bounds and reaches.

\section{Validation of ATLAS analysis}
\label{appn:validation}

The ATLAS analysis \cite{Aaboud:2018xuw} searches for channels obtained from the pair production of toplike vector quarks $t^{\prime}$, one of which decays to a Higgs and a top, and the other decays to a top (or bottom) and any one of the SM bosons ($W^{\pm},H,Z$). The different search channels have been devised to search for different possible final states from these intermediate particles. We focus on the 1-lepton channel described in Tables $1$ and $2$ of \cite{Aaboud:2018xuw}. Preselection is applied by requiring exactly one electron or muon, at least $5$ jets out of which at least $2$ are $b$-tagged, $\slashed{E}_T$ greater than $20$ GeV, and $\slashed{E}_T+m_T^W$ greater than $60$ GeV where $m_T^W$ is the transverse mass of the signal lepton.\\\\
In the analysis top and Higgs candidates are reconstructed by re-clustering signal jets using anti-$k_T$ algorithm with a radius parameter $1.0$. Out of the re-clustered large jets, top candidates are tagged by requiring the $p_T$ to be greater than $300$ GeV, mass greater than $140$ GeV and at least two subjets. Higgs candidates are tagged by requiring $p_T$ to be greater than $200$ GeV, mass between $105$ and $140$ GeV and applying a $p_T$ dependant subjet criteria (exactly two for $p_T$ less than $500$ GeV and one or two for $p_T$ greater than $500$ GeV).\\\\
We mimic the top and Higgs tagging by finding pairs of jets within a distance $\Delta R=1.3$ (defined as $\sqrt{(\Delta y)^2+(\Delta \phi)^2}$, where $y$ is the pseudorapidity and $\phi$ is the azimuthal angle) and satisfying the mass and $p_T$ requirements mentioned previously.\\\\
The different signal regions are chosen for different requirements on the numbers of top tagged large jets, Higgs tagged large jets and $b$-jet multiplicity. We summarize our validation for one signal channel ($t^{\prime}\bar{t^{\prime}}$ with $Br(t^{\prime}\rightarrow Ht)=1$) and one background channel ($t\bar{t}H$) in Table~\ref{tab:validation}.  
\begin{table}[t]
	\centering
	\begin{tabular}{ccccc}
		\hline\hline
		\multirow{2}{*}{Signal Region} &\multicolumn{2}{c}{$t^{\prime}\bar{t^{\prime}},Br(t^{\prime}\rightarrow Ht)=1$} & \multicolumn{2}{c}{$t\bar{t}H$} \\
		& Reported & Our analysis & Reported & Our analysis \\
		\hline
		$\geq2t,0-1H,\geq6j,3b$ & $19.6$ & $15.8$ & $4.9$ & $3$\\
		$1t,0H,\geq6j,\geq4b$ & $21.5$ & $16.6$ & $15$ & $10$\\
		$1t,1H,\geq6j,\geq4b$ & $24.3$ & $11.9$ & $3.8$ & $2$\\
		$\geq2t,0-1H,\geq6j,\geq4b$ & $23.9$ & $20.8$ & $2.8$ & $2$\\
		$\geq0t,\geq2H,\geq6j,\geq4b$ & $14.6$ & $12.7$ & $1.19$ & $3$\\
		\hline\hline
	\end{tabular}
	\caption{\small\it Validation details for the different $1-$lepton signal regions of \cite{Aaboud:2018xuw}. The reported numbers are taken from Table $4$ of the analysis report. We try and match these number for our analysis at an integrated luminosity of $36.1fb^{-1}$.}
	\label{tab:validation}
\end{table}

\section{Single Production Reach}
\label{appn:single_reach}

\begin{figure}[t]
	\centering
	\subfloat[\label{fig:process_single}]{\includegraphics[trim={5cm 20cm 11cm 4cm},clip,scale=0.8]{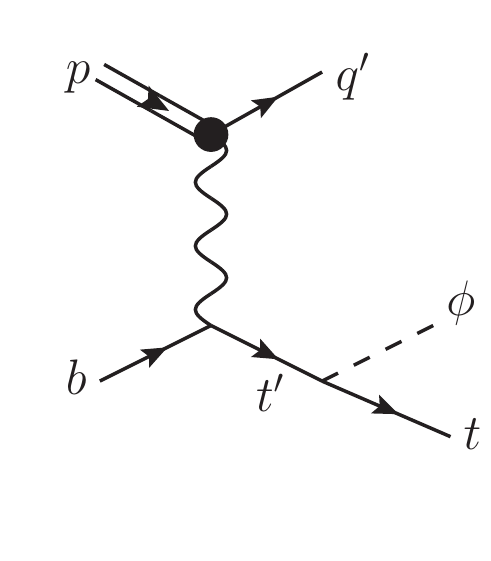}}
	\hspace{2cm}
	\subfloat[\label{fig:reach_single}]{\includegraphics[scale=0.35]{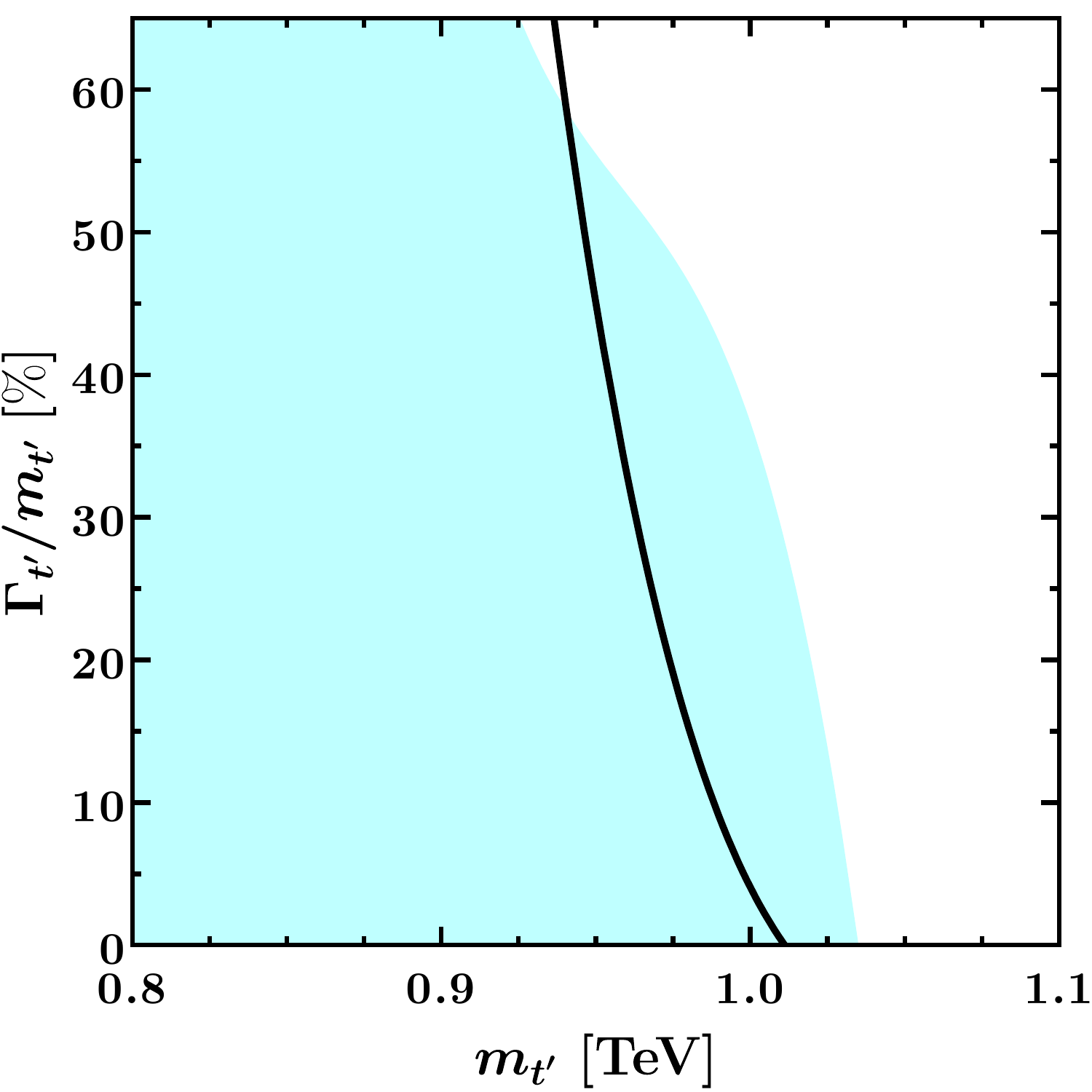}}
	\caption{\small\it (a) Representative feynman diagrams for single production of $t^{\prime}$ and its subsequent major decay. (b) $3\sigma$ HL-LHC for single production of $t^{\prime}$ at $3~ab^{-1}$ integrated luminosity.}
	\label{fig:single}
\end{figure}

We focus on the $pb\rightarrow t^{\prime}\bar{q}\rightarrow\phi t\bar{q}$ channel shown in Fig.~\ref{fig:process_single}. The major SM background contributing to this channel is singletop produced in association with a $W$ boson. The experience with the pair production reach described in Section~\ref{sec:future_strategy} clearly indicates that ML tools provide the most optimistic HL-LHC reach. Based on this we focus on the ML approach using the XGBoost algorithm to study the HL-LHC reach in the single production channel. We vary $m_{t^{\prime}}$ from $0.8$ to $1.3$ TeV in steps of $0.1$ TeV and $\Gamma_{t^{\prime}}/m_{t^{\prime}}$ from $5$ to $60$ in steps of $5$ and generate single production events. Similar to the pair production analysis we mix the signal and background with equal weightage. From the generated data points $80\%$ were used to train the ML algorithm and the rest to test it. We take care to prevent over-training. The obtained HL-LHC $3\sigma$ reach is represented by the black dashed line in Fig.~\ref{fig:reach_single}. As can be seen from the plot even the $3\sigma$ HL-LHC reach is within the current LHC bound obtained in Fig.~\ref{fig:current_bound_gamma_mps100} reaching 1.0 TeV for $\Gamma_{t^{\prime}}/m_{t^{\prime}}=0.05$ because of the low cross section in this channel.

\section{Momentum dependence of top partner mass}
\label{appn:mom_dependence}

\begin{figure}[t]
	\centering
	\includegraphics[scale=0.45]{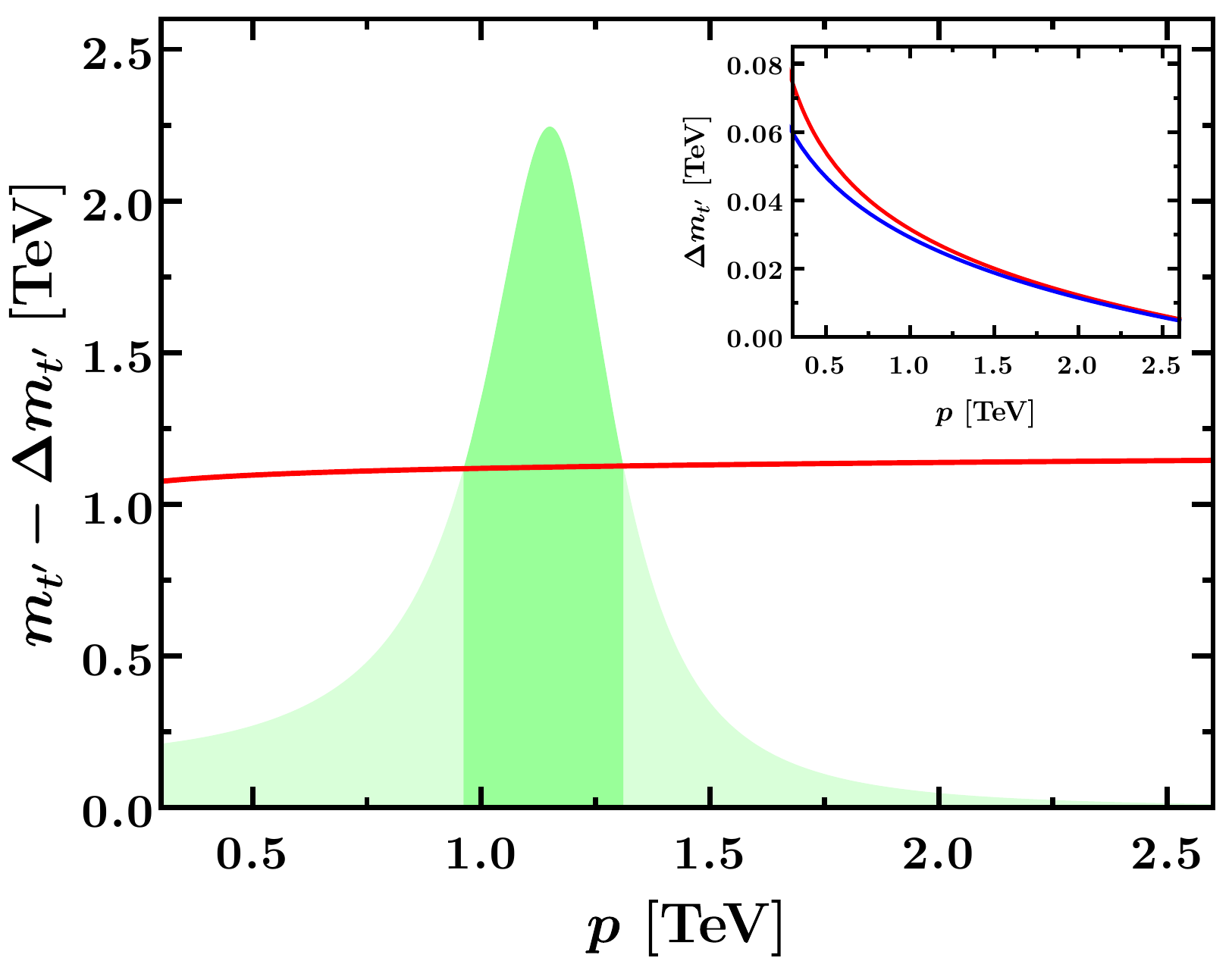}
	\caption{\textit{The red line indicates the momentum dependent top partner mass $(m_{t^\prime} - \Delta m_{t^\prime})$ at $\mu\sim m_{t^{\prime}}$. For reference the green shaded region represents an appropriately normalized BW form of the top partner propagator for $m_{t^\prime} = 1.15$ TeV, $M_\phi = 0.1$ TeV and $\Gamma_{t^\prime}/m_{t^\prime} = 30 \%$. (Inset) The red line indicates the variation of the relative correction $(\Delta m_{t^\prime})$ to the top partner mass. The blue line is a plot of the approximate expression given in Eq. \ref{eq: mass_var_large_p}. } }
	\label{fig:stop_mass_variation}
\end{figure}
In this appendix we demonstrate that the momentum dependent mass correction due to the 1PI contribution to the top partner propagator is numerically insignificant in the region of interest. The $1-$loop self energy diagram given in Figure \ref{fig:top_partner_loop} has a real part that contributes to the top partner mass making it potentially momentum dependent. Taking into account the momentum dependence, the top partner mass can be written as $(m_{t^\prime} - \Delta m_{t^\prime})$ where the leading finite piece in the $\overline{\text{MS}}$ scheme is given by
\begin{equation}
\Delta m_{t^\prime}(p) = \dfrac{m_t g_{\phi}^{\ast 2}}{16 \pi} \int_0^1 dx \ln \dfrac{\mu^2}{\lvert x^2 p^2 - x(p^2 + m_t^2 - M_\phi^2) + m_t^2 \rvert} \ ,
\end{equation}
where $\Delta m_{t^\prime}$ is the momentum dependent contribution to the top partner mass at an energy scale $\mu$. The logarithmic sensitivity to the scale can be read off from the asymptotic $(p \gg m_{t^\prime}, M_\phi)$ behavior of the correction,
\begin{equation}
\Delta m_{t^\prime}(p) \simeq \dfrac{m_t g_{\phi}^{\ast 2}}{16 \pi} \left( 2 + \ln \dfrac{m_{t^\prime}^2}{p^2} \right) .
\label{eq: mass_var_large_p}
\end{equation}
Clearly this logarithmic sensitivity is milder compared to the contribution to the width which remains unprotected from chiral symmetry. As is evident from Figure \ref{fig:stop_mass_variation}, the renormalized mass of the top partner given by $m_{t^\prime} - \Delta m_{t^\prime}$ exhibits soft momentum dependence (variation $\sim 0.8 \%$) within a typical spread of the resonance peak. Consequently, the effect of momentum dependence of the top partner mass has been neglected in our numerical simulations.

\bibliographystyle{JHEP}
\bibliography{Top_partner_1PI.bib}

\end{document}